\documentclass[
  reprint,
  superscriptaddress,
  amsmath,amssymb,
  aps,
  prb,
  floatfix,
  longbibliography,
  showkeys
]{revtex4-2}

\usepackage{graphicx}
\usepackage{dcolumn}
\usepackage{bm}
\usepackage{amsfonts}
\usepackage{booktabs}
\usepackage{multirow}
\usepackage{xcolor}
\usepackage{textcomp}
\usepackage[hidelinks]{hyperref}

\newcommand{\maybeincludegraphics}[2][]{%
  \IfFileExists{#2}{\includegraphics[#1]{#2}}{%
    \fbox{\begin{minipage}[c][0.22\textheight][c]{0.88\linewidth}\centering Missing figure: \texttt{\detokenize{#2}}\end{minipage}}%
  }%
}

\usepackage{xr-hyper}
\usepackage[hidelinks]{hyperref}

\makeatletter
\newcommand*{\addFileDependency}[1]{%
  \typeout{(#1)}%
  \@addtofilelist{#1}%
}
\makeatother

\addFileDependency{PRB_Supplement_Manuscript.tex}
\addFileDependency{PRB_Supplement_Manuscript.aux}

\begin{document}

\title{Field-rigid Ising antiferromagnetism with giant spin-flip fields in Van der Waals UOTe}

\author{Zackary Rehfuss}
\affiliation{Department of Physics, Washington University in St. Louis, St. Louis, Missouri 63130, USA}

\author{Shannon Gould}
\affiliation{Department of Physics, Washington University in St. Louis, St. Louis, Missouri 63130, USA}

\author{Joanna Blawat}
\affiliation{National High Magnetic Field Laboratory, Los Alamos National Laboratory, Los Alamos, New Mexico 87545, USA}

\author{Christopher Broyles}
\affiliation{Department of Physics, Washington University in St. Louis, St. Louis, Missouri 63130, USA}

\author{George Xu}
\affiliation{Department of Physics, Washington University in St. Louis, St. Louis, Missouri 63130, USA}

\author{Yiqing Hao}
\affiliation{Neutron Scattering Division, Oak Ridge National Laboratory, Oak Ridge, TN 37831, USA}

\author{Huibo Cao}
\affiliation{Neutron Scattering Division, Oak Ridge National Laboratory, Oak Ridge, TN 37831, USA}

\author{Thao Dinh}
\affiliation{Department of Chemistry and Chemical Biology, Harvard University, Cambridge, MA 02138, USA}

\author{Suyang Xu}
\affiliation{Department of Chemistry and Chemical Biology, Harvard University, Cambridge, MA 02138, USA}

\author{Dave Graf}
\affiliation{National High Magnetic Field Laboratory, Florida State University, Tallahassee, Florida 32310, USA}

\author{John Singleton}
\affiliation{National High Magnetic Field Laboratory, Los Alamos National Laboratory, Los Alamos, New Mexico 87545, USA}

\author{Sheng Ran}
\affiliation{Department of Physics, Washington University in St. Louis, St. Louis, Missouri 63130, USA}

\date{\today}

\begin{abstract}
Van der Waals antiferromagnets provide a route to thickness-controlled magnetic order, but few combine high-temperature Ising order with conducting, correlated, and topological electronic structure. Here we show that UOTe realizes this combination. 
Magnetic susceptibility reveals a strongly anisotropic paramagnetic response, while neutron diffraction establishes \(c\)-axis antiferromagnetic order below \(T_{\mathrm N}\approx150~\mathrm{K}\) with an order-parameter exponent \(\beta=0.14\), close to the two-dimensional Ising value. Torque magnetometry further shows that the ordered state remains well described by a uniaxial antiferromagnet below the high-field transition. Pulsed-field magnetization up to \(73~\mathrm{T}\) shows that the ordered state survives to very large fields applied along the \(c\) axis before entering a broad metamagnetic regime that begins near \(50~\mathrm{T}\), 
and remains unsaturated at the highest measured field. Angle-dependent proximity detector oscillator measurements show that 
the metamagnetic instability is set by the field component along the ordered moment direction, providing direct evidence for Ising-like field rigidity. 
UOTe therefore establishes a field-rigid Ising antiferromagnet with giant spin-flip fields in a compensated Van der Waals metal, where high-temperature \(c\)-axis order, quasi-two-dimensional magnetic criticality, Kondo-associated uranium \(5f\) hybridization, metallic transport, and symmetry-enabled topology coexist in a single material.
\end{abstract}


\maketitle

\section{Introduction}\label{sec1}

In strictly two-dimensional magnets with short-range exchange and continuous spin symmetry, thermal fluctuations preclude finite-temperature long-range order, as established by the Mermin-Wagner theorem \cite{MERMIN_WAGNER}. This restriction is not violated, but avoided, when magnetic anisotropy breaks continuous spin-rotational symmetry and reduces the relevant symmetry of the ordered state, allowing magnetic order to persist in quasi-two-dimensional materials \cite{2D_ANISOTROPY_REVIEW}. Layered Van der Waals magnets therefore provide a direct materials setting in which dimensionality, magnetic anisotropy, interlayer coupling, spin-orbit interaction, and electronic structure can be tuned against one another and tested through field response \cite{CITE_2DMAG_REVIEW_A,CITE_2DMAG_REVIEW_B,CITE_VDW_MAGNETISM_REVIEW,CITE_VDW_MAGNETISM_ENGINEERING_REVIEW,CITE_SOC_2DMAG_REVIEW,CITE_SPINTRONICS_VDW_HETEROSTRUCTURES,CITE_2DMAG_DETECTION_MODULATION_REVIEW,2D_ANISOTROPY_REVIEW}.


Over the past two decades, a large variety of layered and Van der Waals antiferromagnets have been explored. The transition-metal phosphorus trichalcogenide family, MPX$_3$, is among the best-established examples, which provides mechanically exfoliable, largely insulating layered antiferromagnets where localized $3d$ moments, superexchange, magnetic anisotropy, and field-driven spin reorientation can be studied in a controlled setting \cite{LeFlem1982MPX3,Joy1992MPS3,Samal2021MPX3Review,Kurosawa1983MnPS3FePS3,Lancon2016FePS3,Lee2016FePS3,Wildes2020FePS3,Long2020MnPS3,Kim2019NiPS3,Lancon2018NiPS3,Basnet2021MPS3,Basnet2022MPX3Substitution,Nair2024MnPSe3}. FePS$_3$ is the clearest Ising-like limit in this family, with easy-axis Van der Waals antiferromagnetism that survives to the atomically thin limit and produces a strongly constrained magnetic response \cite{CITE_2D_ISING_ONSAGER,CITE_SIMPLE_MAGNETIC_MODEL_SYSTEMS,CITE_WOLF_ISING_REAL_MATERIALS,CITE_RB2COF4_2D_ISING_AFM,CITE_FEPS3_ATOMICALLY_THIN_ISING_AFM,CITE_ANTIFERROMAGNETISM_REVIEW,CITE_SPINFLOP_UNIAXIAL_AFM}. The broader layered-antiferromagnet landscape now includes weak-anisotropy semiconductors such as CrCl$_3$ and CrPS$_4$, Kitaev-proximate or orbital-active halides such as $\alpha$-RuCl$_3$, VBr$_3$, FeI$_2$, and related trihalides, topological Mn-Bi-Te chalcogenides, rare-earth oxyhalides and tritelluride metals, and layered Dirac systems such as EuMnBi$_2$ \cite{Rahman2021VdWAFM2D,Olsen2024TwoDAFMs,Wang2019CrCl3,Seo2024CrPS4,Zhou2023RuCl3,Hovancik2023VBr3,Katsumata2010FeI2,Yan2019MnBi2Te4,Yan2020MnBi4Te7,Valkov2024MnBi2Te4,Tian2021DyOCl,Volkova2022TbTe3,Akatsuka2024DyTe3,Wan2023GdTe3,May2014EuMnBi2,Masuda2016EuMnBi2,Masuda2018EuMnBi2,Zhu2020EuMnBi2}. These systems show that layered antiferromagnetism can couple magnetic anisotropy and field-driven spin textures to itinerant carriers, orbital character, lattice symmetry, topology, and collective electronic reconstruction.

Uranium compounds add the further complexity of spatially extended $5f$ orbitals, where local-moment magnetism, spin-orbit coupling, intra-atomic correlations, ligand hybridization, and itinerant electronic structure can coexist within the same material. While these intertwined effects have been extensively studied in three-dimensional uranium magnets\cite{Leciejewicz1967USb2,Wawryk2006USb2,Giannakis2019USb2,Chen2019USb2,Stillwell2017USb2,Sandratskii2019USb2,USB2,UBITE,U3Bi4Ni3,U2Fe3Ge}, layered uranium magnetic systems have only recently begun to attract broader attention\cite{UOTE_MAIN_PAPER,UOTE_NEW_PAPER,UTE3_JUSTIN,UTE3_LANL}. 

\begin{figure}[t]
\centering
\maybeincludegraphics[width=\linewidth,height=0.34\textheight,keepaspectratio]{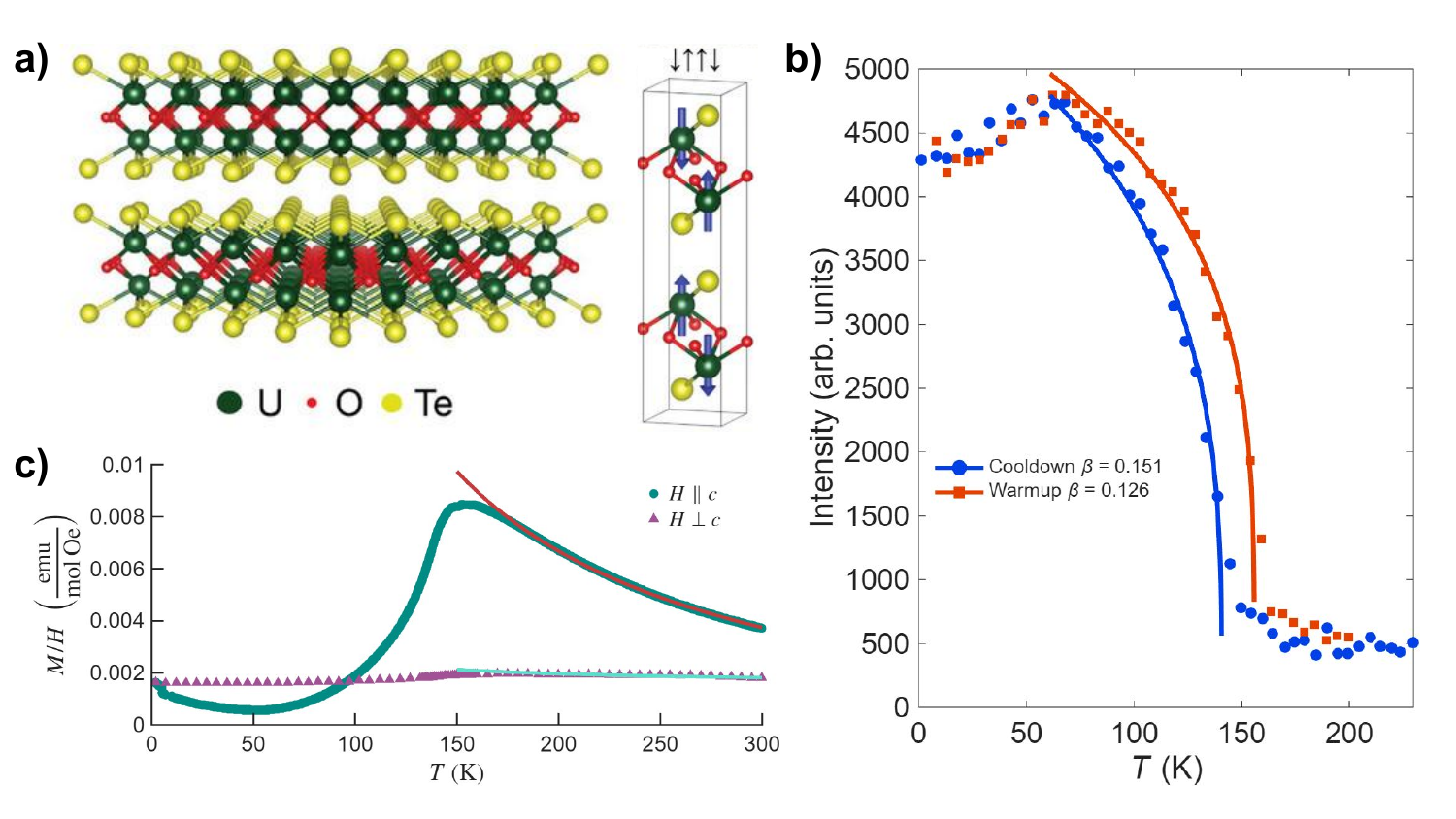}
\caption{\textbf{a)} Crystal and magnetic structure of UOTe. Left, cross-sectional view of the layered structure, showing the Te-terminated Van der Waals gaps between neighboring structural units. Right, magnetic structure showing the doubled magnetic unit cell with the $\downarrow\uparrow\uparrow\downarrow$ arrangement. \textbf{b)} Temperature dependence of the magnetic Bragg peak intensity at $\mathbf{Q}=(1,0,0.5)$. Data were collected in 5 K steps, with each point integrated for 1 minute after a 1 minute equilibration period. Solid lines are fits to a critical scaling form, and the extracted values of $\beta$ are listed in the legend. The fitted values are close to the two-dimensional Ising limit. \textbf{c)} Temperature dependence of the magnetic susceptibility measured in an applied field of 1000 Oe for $H \parallel c$ and $H \parallel ab$. Solid lines show Curie--Weiss fits to the high-temperature response. A strong magnetic anisotropy is observed between the two field orientations.}
\label{fig:intro_full}
\end{figure}

UOTe is one of the first uranium-based Van der Waals compounds. It orders antiferromagnetically near $T_{\mathrm{N}}\approx150~\mathrm{K}$ in the layered Te-U-O-U-Te lattice shown in Fig.~\ref{fig:intro_full}a, with uranium moments aligned along the crystallographic $c$ axis in a fully compensated $\downarrow\uparrow\uparrow\downarrow$ magnetic structure \cite{UOTE_MAIN_PAPER}. ARPES and DFT+$U$ identify UOTe as an antiferromagnetic Dirac semimetal, while resonant photoemission reveals a temperature-dependent U $5f$ flat band and hybridization with Dirac-like conduction states, establishing the coexistence of antiferromagnetism, topology, and Kondo interaction within the same electronic structure \cite{UOTE_MAIN_PAPER}. Its magnetic structure preserves combined parity and time-reversal symmetry, $\mathcal{P}\mathcal{T}$, in the bulk and odd-layer limit, but breaks this symmetry in even layers; bilayer calculations predict a fully compensated antiferromagnetic Chern-insulating state with nonzero Chern number despite zero net magnetization \cite{UOTE_MAIN_PAPER}. UOTe therefore provides a setting in which strongly anisotropic uranium $5f$ antiferromagnetism can be tested in a correlated, topological Van der Waals lattice.

Here we show that this ordered state is unusually rigid against magnetic field. Susceptibility and neutron diffraction establish a strongly anisotropic antiferromagnet with $c$-axis moments and critical behavior close to the two-dimensional Ising limit. Torque magnetometry shows that the low-field state behaves as a uniaxial antiferromagnet, while pulsed-field magnetization and angle-dependent proximity detector oscillator measurements show that the high-field instability occurs only at large fields and is governed by the $c$-axis component of the applied field. Instead of a single conventional spin flip, UOTe enters a broad, hysteretic, and internally structured metamagnetic regime. 

\section{Methods}\label{sec2}

Single crystals of UOTe were synthesized by chemical vapor transport in a two-zone furnace. High-purity tellurium (\(99.9999\%\), Alfa Aesar) and depleted uranium metal (New Brunswick Laboratory) were combined in an equimolar ratio and loaded into an evacuated quartz ampoule together with iodine as the transport agent (\(1~\mathrm{mg~cm^{-3}}\)). The sealed tube was placed across a temperature gradient of \(1030~^{\circ}\mathrm{C}\) at the source and \(970~^{\circ}\mathrm{C}\) at the sink for seven days. After the transport reaction, the furnace was cooled to ambient temperature, yielding shiny, plate-like UOTe single crystals suitable for physical property measurements.

Magnetization measurements were carried out using a vibrating sample magnetometer (VSM). A magnetic field of \(1000~\mathrm{Oe}\) was applied during data acquisition. The temperature was varied from \(2~\mathrm{K}\) to \(300~\mathrm{K}\) for fields oriented within the \(ab\) plane and from \(2~\mathrm{K}\) to \(400~\mathrm{K}\) for fields applied along the crystallographic \(c\) axis.

Torque magnetometry measurements were conducted at the High Magnetic Field Laboratory in Tallahassee, Florida, in fields up to \(41.5~\mathrm{T}\) and temperatures down to \(350~\mathrm{mK}\). Crystals with a nominal thickness below \(100~\mu\mathrm{m}\) and approximate lateral dimensions of \(200~\mu\mathrm{m} \times 200~\mu\mathrm{m}\) were selected under optical inspection. Each crystal was fixed with GE varnish onto a piezoresistive torque cantilever and measured in magnetic fields up to \(41.5~\mathrm{T}\) at a base temperature of \(350~\mathrm{mK}\).


Single-crystal neutron diffraction measurements were performed on the HB-3A DEMAND four-circle diffractometer at the High Flux Isotope Reactor, Oak Ridge National Laboratory \cite{Chakoumakos2011HB3A,Cao2019DEMAND}. A Si~(220) monochromator was used to select a neutron wavelength of \(1.542~\text{\AA}\). Several UOTe single crystals were coaligned on a Si substrate to form a composite sample with a total mass of approximately \(0.1~\mathrm{g}\) and a mosaic spread of \(\pm5^{\circ}\). The sample assembly was mounted on the four-circle goniometer and cooled to \(5~\mathrm{K}\) using a closed-cycle helium refrigerator. Data reduction was performed using the \textsc{ReTIA} software package \cite{Hao2023ReTIA}. Symmetry analysis was carried out using the \textsc{Bilbao Crystallographic Server} \cite{Aroyo2006Bilbao}, and magnetic-structure refinements were performed with the \textsc{FullProf} suite \cite{RodriguezCarvajal1993FullProf}.

Pulsed-field magnetization and proximity detector oscillator (PDO) measurements were carried out at the Pulsed Field Facility of the National High Magnetic Field Laboratory at Los Alamos National Laboratory in magnetic fields up to 73~T and temperatures down to 570~mK. Magnetization was measured inductively using a compensated pick-up coil magnetometer with the field applied parallel to the crystallographic \(c\) axis. In this technique, the coil voltage is proportional to \(dM/dt\) and was numerically integrated to obtain \(M(H)\) after subtraction of the background signal measured under identical conditions without the sample. PDO measurements were performed using plate-like single crystals with their large faces spanning the \(ab\) plane, mounted on the detection coil inside a rotator. The sample was rotated from \(H \parallel c\) toward \(H \parallel ab\) to determine the angular evolution of the field-induced transition.


\section{Results}\label{sec3}


\subsection{Magnetization and Curie--Weiss Analysis}\label{subsec_mag_CW}

We first establish the magnetic anisotropy from the temperature-dependent susceptibility for fields applied parallel to the crystallographic $c$ axis and within the $ab$ plane. As shown in Fig.~\ref{fig:intro_full}c, the response is strongly anisotropic over the full measured temperature range, with $\chi(T)$ for $H \parallel c$ substantially larger than for $H \parallel ab$. Both field orientations show a clear anomaly at $T_{\mathrm N}$, marking the onset of antiferromagnetic order. The much larger response for $H \parallel c$ identifies the crystallographic $c$ axis as the magnetic easy axis already in the paramagnetic regime.

To quantify the paramagnetic response above $T_{\mathrm N}$, we fit the susceptibility to a Curie--Weiss form with a temperature-independent background,
\[
\chi(T)=\chi_{0}+\frac{C}{T-\theta_{\mathrm{CW}}},
\]
where $\chi_{0}$ is the temperature-independent contribution, $C$ is the Curie constant containing the effective moment $\mu_{\mathrm{eff}}$, and $\theta_{\mathrm{CW}}$ is the Curie--Weiss temperature. 
For $H \parallel c$, the fit gives $\theta_{\mathrm{CW}}^{c}=-30~\mathrm{K}$, $\mu_{\mathrm{eff}}^{c}=3.2~\mu_{\mathrm{B}}$, and $\chi_{0}^{c}=-2.2\times10^{-3}$. For $H \parallel ab$, the Curie--Weiss temperature was fixed to the value obtained for $H \parallel c$, giving $\theta_{\mathrm{CW}}^{ab}=-30~\mathrm{K}$, $\mu_{\mathrm{eff}}^{ab}=0.64~\mu_{\mathrm{B}}$, and $\chi_{0}^{ab}=1.7\times10^{-3}$.

The anisotropy of the fitted Curie constants can be expressed as an effective anisotropic $g$-factor ratio using \(C_i \propto g_i^2\) for a fixed moment manifold:
\[
\frac{g_c}{g_{ab}}=\sqrt{\frac{C_c}{C_{ab}}}=4.97.
\]

As a dimensionless measure of the uniaxial character of the paramagnetic response, "Ising-ness" can be defined as
\[
\mathcal{A}_{g}
=
\frac{\left(g_c/g_{ab}\right)^2-1}
{\left(g_c/g_{ab}\right)^2+1}
=
\frac{C_c-C_{ab}}{C_c+C_{ab}}.
\]
This quantity is zero for an isotropic Curie response and approaches unity when the transverse Curie response vanishes. For UOTe, the fitted Curie constants give $\mathcal{A}_{g}=0.922$, indicating that strong anisotropy is already present above $T_{\mathrm N}$. 

\begin{figure}[t]
    \centering
    \maybeincludegraphics[width=\linewidth]{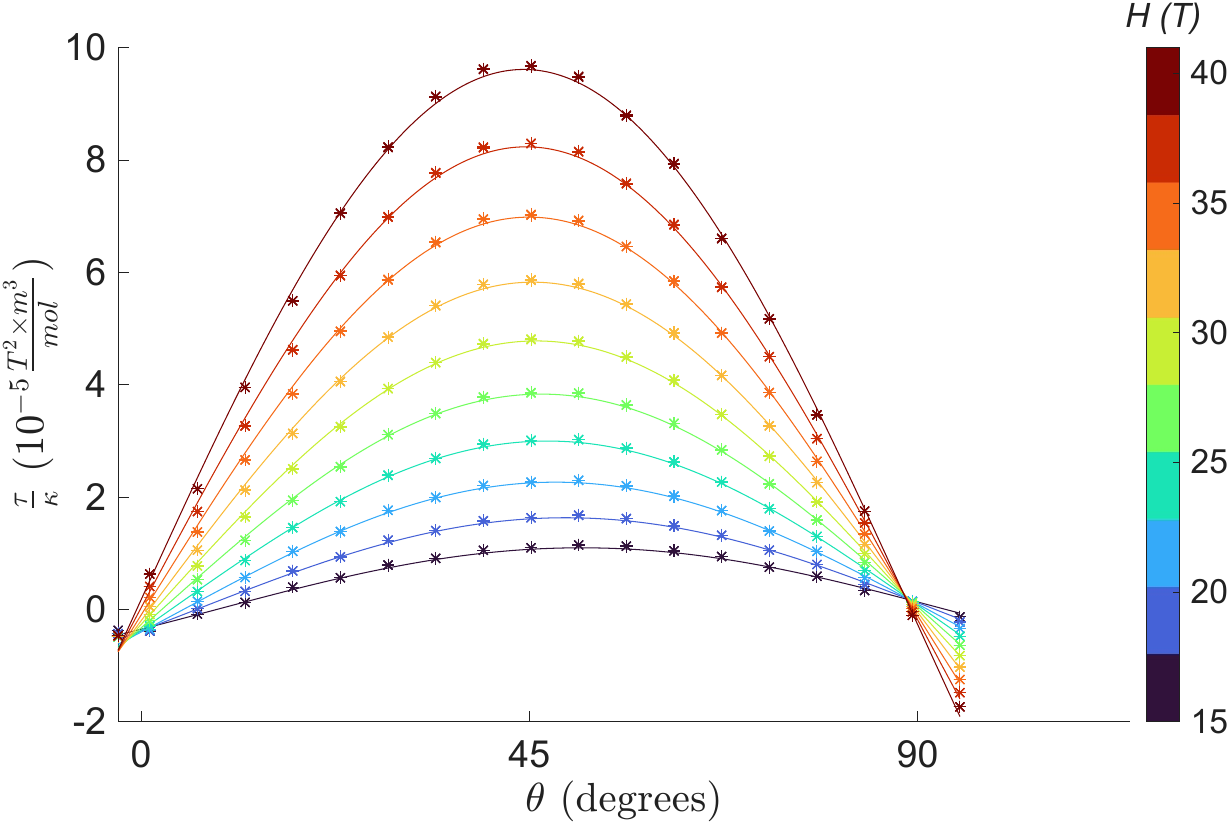}
    \caption{\textbf{Angular cuts through the torque data.} 
    Representative angular cuts extracted from the torque data at fixed magnetic fields. Solid lines are fits to the \(\sin(2\theta)\) angular dependence expected for a uniaxial two-sublattice antiferromagnet, \(\tau \propto \Delta\chi H^2\sin(2\theta)\). The agreement shows that the angular response remains consistent with a simple uniaxial antiferromagnetic state throughout the measured field range.}
    \label{fig:Torque_Fit_AngleSlices}
\end{figure}

\subsection{Neutron Diffraction and Critical Behavior}\label{subsec_neutron}

Having established the strong easy-axis susceptibility, we next track the development of antiferromagnetic order directly through neutron diffraction. The temperature dependence of the magnetic Bragg peak intensity at $\mathbf{Q}=(1,0,0.5)$ is shown in Fig.~\ref{fig:intro_full}b. The intensity increases continuously on cooling below $T_{\mathrm N}$, consistent with the onset of long-range antiferromagnetic order. Near the transition, the data were fit to the standard order-parameter form
\[
I(T)\propto \left(1-\frac{T}{T_{\mathrm N}}\right)^{2\beta},
\]
where the Bragg intensity is proportional to the square of the ordered moment.

Fitting the critical region from $61$ to $T_{\mathrm N}=148$~K gives a best-fit exponent of $\beta=0.1385$. As a stability check, we averaged fits over lower-temperature cutoffs from $31$ to $81$~K, giving $\beta=0.1278$ and $T_{\mathrm N}=147.7$~K. The cutoff range extends up to $81$~K because the neutron intensity is nearly saturated below this temperature, whereas the rapid growth toward $T_{\mathrm N}$ occurs above it. Details of the fitting procedure are provided in the Supplemental Material. The extracted exponent is close to the two-dimensional Ising value of $1/8$, indicating quasi-two-dimensional Ising-like critical behavior.

\begin{figure*}[t]
\centering
\maybeincludegraphics[
  width=0.95\textwidth,
  height=0.68\textheight,
  keepaspectratio
]{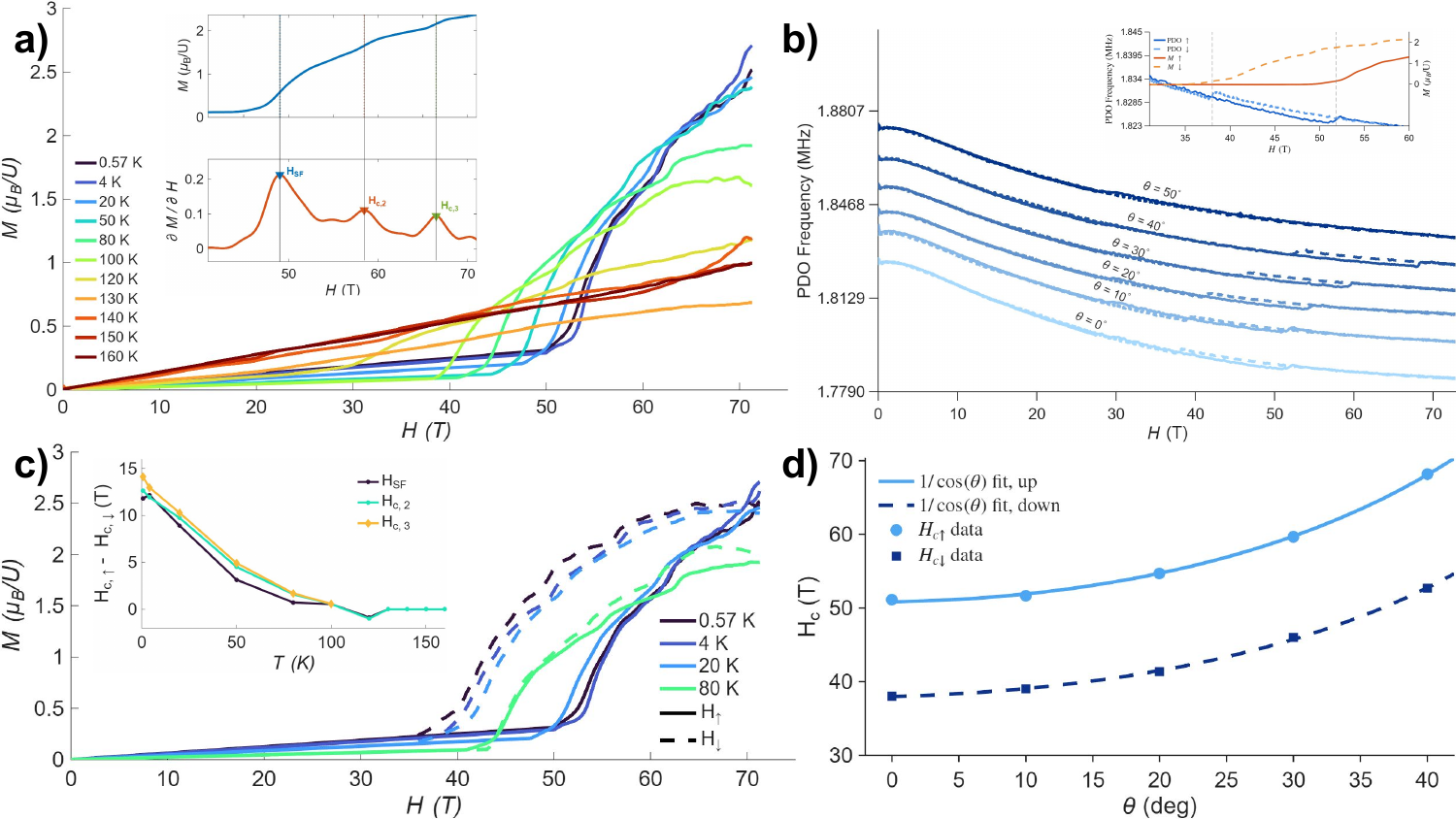}
\caption{\textbf{a)} Magnetization of UOTe as a function of field for temperatures from 570 mK to 160 K, measured in pulsed fields up to 73 T. The data were normalized using vibrating sample magnetometer measurements from earlier experiments. The inset shows an expanded view of the spin-flip transition and its field derivative, highlighting the three peaks used to track the phase boundary. \textbf{b)} PDO frequency shift as a function of field at 570 mK for field angles measured with respect to the $c$ axis. A clear transition is observed, and the inset compares the PDO response with the magnetization data near the onset of the spin-flip transition. The critical field increases with angle from $0^\circ$ to $50^\circ$, beyond which the transition is not resolved up to 73 T. \textbf{c)} Up-sweep and down-sweep magnetization at selected temperatures from 570 mK to 80 K, showing the evolution and closing of the hysteresis loop with increasing temperature. The inset shows the hysteresis width extracted from the derivative peak positions on up-sweep and down-sweep traces for the three tracked features. \textbf{d)} Angular dependence of the critical field determined from PDO measurements on up-sweep and down-sweep for $H \parallel c$. Solid lines are fits to a $1/\cos(\theta)$ form, indicating scaling with the $c$-axis component of the applied field.}
\label{fig:mag_pdo_full}
\end{figure*}

\subsection{Torque Magnetometry}\label{subsec_torque}

To test whether the ordered state retains a simple uniaxial response below the high-field transition, we measured the magnetic torque as a function of field and angle. 
Across the full accessible field range up to 41 T, the data are well described by the angular dependence expected for a uniaxial two-sublattice antiferromagnet\cite{Kittel1951AFMR,Nagamiya1955Antiferromagnetism},
\[
\tau \propto \Delta\chi\,H^{2}\sin(2\theta),
\]
where $\Delta\chi=\chi_c-\chi_{ab}$ is the susceptibility anisotropy. The torque amplitude therefore tracks the difference between the $c$-axis and in-plane magnetic responses. Representative angular cuts in Fig.~\ref{fig:Torque_Fit_AngleSlices} follow the same form across the measured field range.

No additional anomalies are resolved in the torque response up to 41~T. Instead, the signal remains smooth and retains the same uniaxial angular dependence throughout, consistent with a stable easy-axis antiferromagnetic state below the onset of the high-field metamagnetic regime.

\subsection{Pulsed-Field Magnetization}\label{subsec_pulsedM}

Having established that the ordered state remains uniaxial up to 41~T, we next probe the field range where the antiferromagnetic state is destabilized. Figure~\ref{fig:mag_pdo_full}a shows the pulsed-field magnetization for $H \parallel c$ from 570~mK to 160~K. From 570~mK to 120~K, $M(H)$ remains only weakly field dependent over a broad low-field interval, consistent with the rigid easy-axis response inferred from susceptibility, neutron diffraction, and torque magnetometry. At higher fields, the magnetization enters a pronounced metamagnetic regime. At base temperature, the onset occurs near 50~T, but the transition does not appear as a single sharp spin-flip step. Instead, $M(H)$ evolves through a broad sequence of rapid increases and intermediate shoulders. The high-field evolution remains incomplete at 73~T, showing that the field-driven reconstruction extends over more than 20~T and does not reach full saturation within the available field window. 

\begin{figure*}[t]
\centering
\maybeincludegraphics[width=0.95\textwidth,height=0.40\textheight,keepaspectratio]{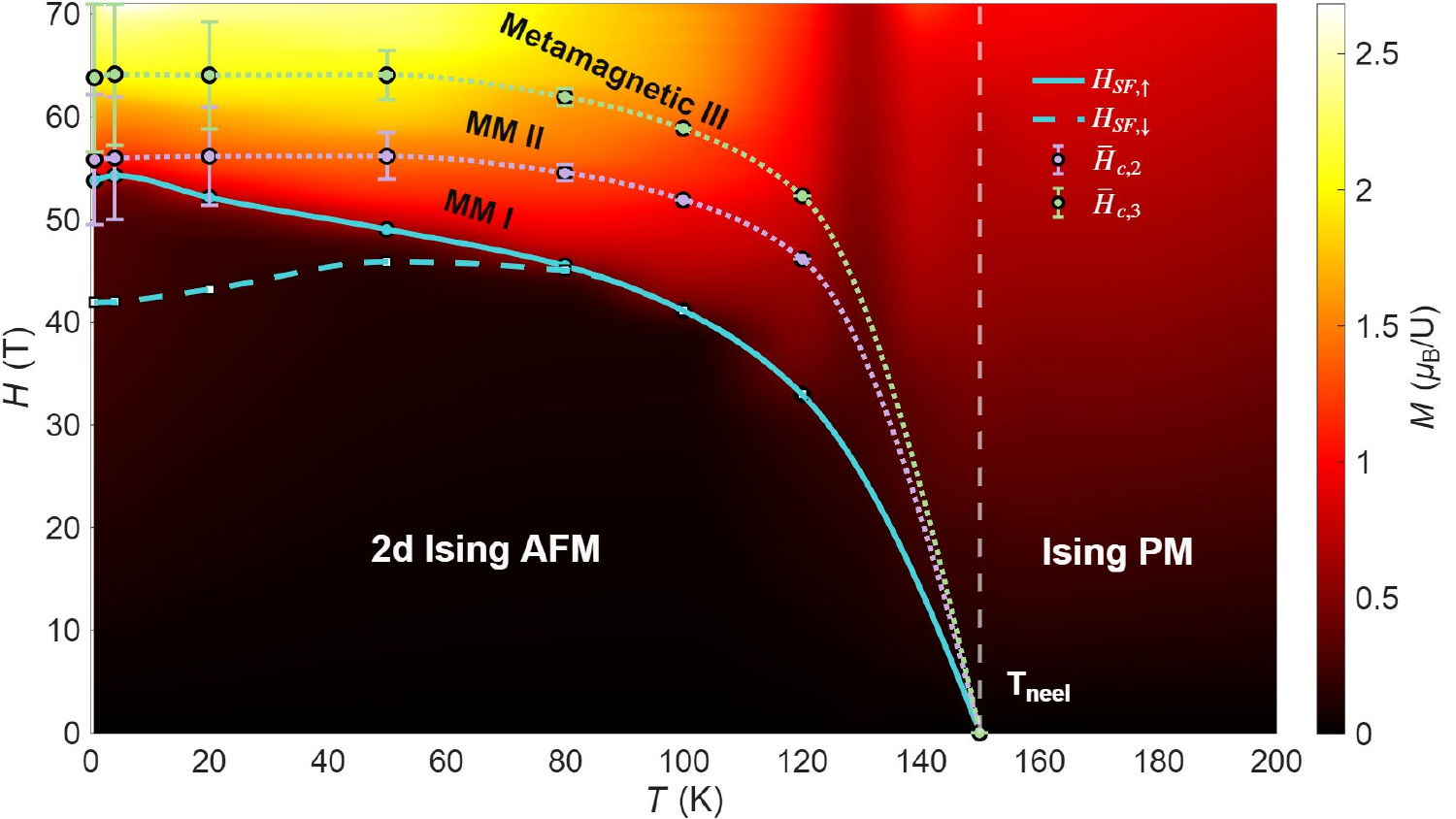}
\caption{ Magnetic field-temperature phase diagram of UOTe constructed from pulsed-field magnetization and PDO measurements, together with neutron scattering, magnetic susceptibility, and torque magnetometry. Above $T_{\mathrm{N}} \approx 150$ K, UOTe is in the paramagnetic state. Below $T_{\mathrm{N}}$, at least three field-induced branches emerge from the metamagnetic transition region and define the evolution of the ordered state with temperature and field. The blue boundary marks the spin-flip transition, with separate up-sweep and down-sweep values shown where hysteresis is resolved. The green and pink branches correspond to the higher-field derivative features identified from the magnetization data. Vertical bars denote the up-sweep and down-sweep values, and solid lines indicate the average field position of each feature. The regions labeled MM I and MM II denote distinct metamagnetic regimes separated by these field scales.}
\label{fig:phase_diagram}
\end{figure*}

The internal structure of this metamagnetic regime is resolved more clearly in the field derivative, shown in the inset of Fig.~\ref{fig:mag_pdo_full}a. Three distinct peaks appear in $dM/dH$, defining characteristic fields associated with the onset, intermediate evolution, and upper portion of the transition. These features shift systematically with temperature and provide the field scales used to construct the phase diagram in Fig.~\ref{fig:phase_diagram}. Between 120~K and 160~K, the response crosses over toward a more nearly linear susceptibility, consistent with the loss of the low-temperature metamagnetic structure as the system approaches and exceeds the antiferromagnetic transition.

A second important feature is the pronounced hysteresis between up-sweeps and down-sweeps at low temperature, shown in Fig.~\ref{fig:mag_pdo_full}c. We define the hysteresis width of each derivative feature as
\[
\Delta H_i(T)=H_i^{\uparrow}(T)-H_i^{\downarrow}(T),
\]
where $H_i^{\uparrow}$ and $H_i^{\downarrow}$ are the corresponding peak positions in $dM/dH$ on the up-sweep and down-sweep, respectively. At base temperature, the overall hysteretic width across the metamagnetic region is approximately 14~T. As shown in the inset of Fig.~\ref{fig:mag_pdo_full}c, the hysteresis decreases steadily with increasing temperature and disappears near 100~K.

Taken together, the pulsed-field magnetization shows that the antiferromagnetic state of UOTe survives to exceptionally large $c$ axis fields before reconstructing through a broad, hysteretic, and multi-stage metamagnetic regime. This behavior is distinct from a conventional narrow spin-flip transition and establishes that the high-field state is not fully saturated by 73~T.

\subsection{Angle-Dependent PDO Measurements}\label{subsec_pdo}

The angle-dependent PDO measurements provide a direct test of whether the high-field instability is controlled by the total magnetic field or by the field component along the Ising axis. Figure~\ref{fig:mag_pdo_full}b shows the PDO frequency shift at 570~mK for fields rotated away from $H \parallel c$. The inset compares the PDO response for $H \parallel c$ with the magnetization measured in the same field orientation, showing that the PDO anomalies coincide with the onset and completion of the high-field spin-flip process identified in $M(H)$. As the field is tilted toward the $ab$ plane, the anomaly shifts systematically to higher total field while retaining the same overall form. This behavior is the defining expectation for an easy-axis instability governed by the $c$ axis field projection.

The extracted critical fields are summarized in Fig.~\ref{fig:mag_pdo_full}d. Both up-sweep and down-sweep values follow
\[
H_{\mathrm{c}}(\theta)=\frac{H_{\mathrm{c}}(0)}{\cos\theta},
\]
so that
\[
H_{\mathrm{c}}(\theta)\cos\theta \approx H_{\mathrm{c}}(0).
\]
Thus, the transition occurs when the $c$-axis component of the applied field reaches the same critical value, independent of the transverse field component within the measured angular range.
No separate anomaly associated with the in-plane field component is resolved. As the field is rotated farther toward $H \parallel ab$, the required total field increases according to the $1/\cos\theta$ form and the transition moves beyond the 73~T field window above approximately $50^\circ$.
Together, the angle-resolved PDO data show that the high-field transition is locked to the easy axis. This uniaxial-field scaling is a central signature of Ising-like behavior. 

\begin{figure*}[t]
\centering
\maybeincludegraphics[
  width=0.95\textwidth,
  height=0.50\textheight,
  keepaspectratio
]{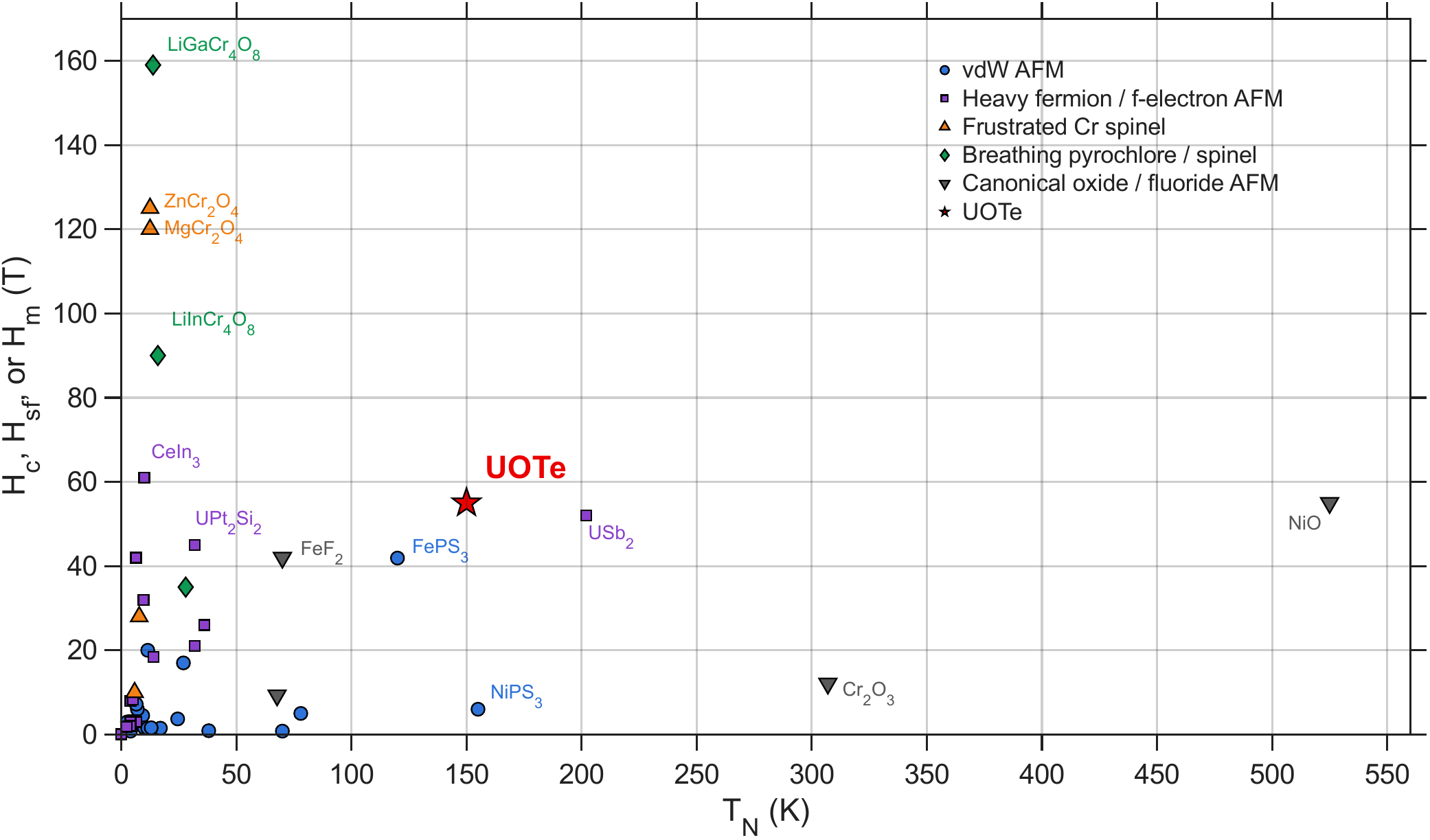}
\caption{Critical field scale plotted against N\'eel temperature for a representative set of antiferromagnetic materials. The vertical axis denotes the reported spin-flip field, spin-flop field, metamagnetic field, or other relevant critical field, labeled here as $H_c$, $H_{\mathrm{sf}}$, or $H_m$. The comparison includes compounds from Van der Waals antiferromagnets, heavy-fermion and other $f$-electron antiferromagnets, frustrated chromium spinels, breathing pyrochlore or spinel antiferromagnets, and canonical oxide or fluoride antiferromagnets. Literature values used to construct the comparison are taken from Refs.~\cite{Wildes2020FePS3,Long2020MnPS3,Nair2024MnPSe3,Basnet2021MPS3,Wang2019CrCl3,Tian2021DyOCl,Hovancik2022UI3,Volkova2022TbTe3,Akatsuka2024DyTe3,Yumigeta2024RTe3Alloying,Wan2023GdTe3,Katsumata2010FeI2,Valkov2024MnBi2Te4,Yan2020MnBi4Te7,Hovancik2023VBr3,Seo2024CrPS4,Zhou2023RuCl3,Aoki2013HeavyFermionsHighField,Stillwell2017USb2,Miyata2013CdCr2O4HighField,Ueda2006HgCr2O4,Miyata2011ZnCr2O4PRL,Miyata2014MgCr2O4,Okamoto2013LiGaLiInCr4O8,Okamoto2017LiInCr4O8HighField,Gen2023LiGaCr4O8PNAS,Okamoto2018CuInCr4S8,Gen2020CuInCr4S8,Felcher1996MnF2SpinFlop,Jacobs1961MnF2SpinFlop,Li2019FeF2SpinSeebeck,Foner1963Cr2O3AFMR,Machado2017NiOSpinFlop}.}
\label{fig:other_compounds}
\end{figure*}

\section{Discussion}\label{sec:discussion}


The results establish UOTe as a strongly anisotropic antiferromagnet whose magnetic rigidity persists from the paramagnetic regime into the high-field metamagnetic state. Above $T_{\mathrm N}$, the susceptibility is already strongly biased toward the crystallographic $c$ axis, with a Curie--Weiss anisotropy corresponding to \(g_c/g_{ab}=4.97\) and \(\mathcal{A}_{g}=0.922\). This value is already close to the Ising limit of \(\mathcal{A}_{g}=1\). For comparison, CoNb$_2$O$_6$, a canonical finite-transverse Ising-chain material, has reported EPR principal values \(g_x=3.33\), \(g_y=3.01\), and \(g_z=6.39\), corresponding to \(\mathcal{A}_g\simeq0.60\) when the transverse response is averaged over the two smaller principal axes \cite{Ringler2022CoNb2O6SingleIon}. Even FePS$_3$, the prototype Van der Waals Ising antiferromagnet, is better described by a large anisotropy energy and an anisotropic torque response than by a nearly extinguished transverse Curie response; its transverse-field behavior involves spin tilting and strong magnetolattice coupling rather than a simple locked-moment limit \cite{Nauman2021FePS3Anisotropy,Wildes2020FePS3}. Stronger finite-transverse anisotropy is found in rare-earth and heavy-fermion systems: YbAlO$_3$ has \(g_{zz}=7.6\) and \(g_{xx}\simeq g_{yy}=0.46\), giving \(g_{zz}/g_{xx}=16.5\) and \(\mathcal{A}_g=0.993\), while the Kondo lattice YbRh$_2$Si$_2$ has \(g_{\perp}=3.56\) and \(g_{\parallel}=0.17\), giving \(g_{\perp}/g_{\parallel}=20.9\) and \(|\mathcal{A}_g|=0.995\), although its anisotropy is easy-plane rather than easy-axis \cite{Wu2019YbAlO3,Sichelschmidt2003YbRh2Si2ESR}. A uranium heavy-fermion upper comparison is URu$_2$Si$_2$, where hybridized quasiparticles show \(g_c/g_{ab}>30\), corresponding to \(\mathcal{A}_g>0.998\) \cite{Altarawneh2012URu2Si2Anisotropy,Chandra2013URu2Si2Hastatic}. UOTe therefore does not reach the nearly locked rare-earth or URu$_2$Si$_2$ limit, but \(\mathcal{A}_g=0.922\) places it much closer to that limit than to conventional finite-\(g_\perp\) Ising-chain or vdW Ising-antiferromagnet comparisons, while retaining metallic uranium \(5f\) electronic structure, Kondo-related hybridization, and topological bands.

This anisotropy is inherited by the ordered state. Neutron diffraction shows that the ordered moments align along $c$, while the magnetic Bragg intensity follows critical behavior close to the two-dimensional Ising value. Susceptibility therefore identifies the easy axis before long-range order develops, and neutron diffraction shows that the transition itself has reduced-dimensional Ising-like character. Torque magnetometry then confirms that the ordered state remains in this same uniaxial regime up to 41~T, with no resolved low-field anomaly or change in angular form.

The angle-dependent PDO measurements provide the most direct high-field test of this Ising character. The metamagnetic anomaly follows \(H_{\mathrm c}(\theta)=H_{\mathrm c}(0)/\cos\theta\), so the transition occurs at an approximately $c$ axis field component \(H\cos\theta\). This scaling rules out a response controlled simply by the total applied field and shows that the instability couples primarily to the field component conjugate to the $c$-axis moment. 
The high-field response is therefore not merely anisotropic; it is locked to the Ising axis.

Once this field scale is reached, UOTe does not undergo a single sharp spin-flip transition. Instead, pulsed-field magnetization reveals a broad metamagnetic regime with multiple resolved maxima in \(dM/dH\), strong low-temperature hysteresis, and no full saturation by 73~T. These features are not consistent with a simple two-state reversal of rigid local moments. The three derivative peaks define internal field scales within the transition, while the absence of a high-field plateau shows that the reconstruction remains incomplete in the available field window. A minimal interpretation is that field sequentially destabilizes several closely competing collinear or nearly collinear configurations. In a layered \(\downarrow\uparrow\uparrow\downarrow\) antiferromagnet, these could involve inequivalent layer reversals, domain-dependent pathways, partially polarized layer-resolved states, or coupled magnetoelastic and electronic reconstructions. The angular PDO data argue against a distinct transverse-field-driven canting instability in the measured field range, so interpretations based primarily on ordinary spin canting are disfavored.

The temperature dependence of the metamagnetism links this high-field reconstruction to the correlated electronic background. Above roughly 100~K, the high-field features broaden and no hysteresis is resolved. Below this scale, the derivative structure sharpens and a large hysteresis develops, reaching an overall width of about 14~T at 570~mK. Prior work on UOTe identified a temperature-dependent U \(5f\) flat band that emerges below about 100~K and hybridizes with Dirac-like conduction states \cite{UOTE_MAIN_PAPER}. The coincidence of these temperature scales indicate Kondo hybridization might play a role in the hysteretic metamagnetism. Further investigation is required to fully understand. 


The comparison in Fig.~\ref{fig:other_compounds} places the field scale in a broader antiferromagnetic context. For each compound, we compare the reported critical field scale, whether labeled as \(H_c\), \(H_{\mathrm{sf}}\), or \(H_m\), with \(T_{\mathrm N}\). Using \(g=2\) as a common qualitative estimate gives \(g\mu_B H \simeq 0.116~\mathrm{meV/T}\times H\), while \(k_B T_{\mathrm N}\simeq 0.086~\mathrm{meV/K}\times T_{\mathrm N}\). For UOTe, \(g\mu_B H_{\mathrm{sf}}\simeq 6.4~\mathrm{meV}\) and \(k_B T_{\mathrm N}\simeq 12.9~\mathrm{meV}\), giving \(g\mu_B H_{\mathrm{sf}}/k_B T_{\mathrm N}\sim0.5\). FePS$_3$ is the closest localized Van der Waals benchmark because \(T_{\mathrm N}\sim120~\mathrm{K}\) and high-field transitions near \(40~\mathrm{T}\) give a comparable ratio, but FePS$_3$ is a localized insulating \(3d\) magnet rather than a conducting \(5f\) system \cite{Wildes2020FePS3,Lee2016FePS3,Lee2023FePS3GiantAnisotropy}. Most other Van der Waals antiferromagnets in the comparison have either much smaller critical fields or much lower ordering temperatures, so they do not combine high \(T_{\mathrm N}\), large spin flip field scale, and electronic itinerancy in the same way.

The broader materials comparison separates UOTe from the other high-field families. In frustrated chromium spinels and breathing pyrochlore or spinel antiferromagnets, the largest critical fields can exceed \(100~\mathrm{T}\), giving \(g\mu_B H_c \gg k_B T_{\mathrm N}\). These systems occupy the high-field, low-\(T_{\mathrm N}\) region of Fig.~\ref{fig:other_compounds}, where large fields act on comparatively low ordering-temperature states and often reflect frustration, spin-lattice coupling, and field-induced selection among nearly competing configurations \cite{Miyata2013CdCr2O4HighField,Ueda2006HgCr2O4,Miyata2011ZnCr2O4PRL,Miyata2014MgCr2O4,Okamoto2017LiInCr4O8HighField,Gen2023LiGaCr4O8PNAS,Gen2020CuInCr4S8}. At the opposite extreme, canonical antiferromagnetic oxides such as NiO and Cr$_2$O$_3$ have very large ordering temperatures, so \(g\mu_B H_c \ll k_B T_{\mathrm N}\), but they are insulating three-dimensional oxides~\cite{Foner1963Cr2O3AFMR,Machado2017NiOSpinFlop}. Among the \(f\)-electron systems, USb$_2$ is the closest comparison in combined ordering and field scale, with \(T_{\mathrm N}\approx202~\mathrm{K}\) and a high-field transition near \(52~\mathrm{T}\), but it is not a Van der Waals material~\cite{Stillwell2017USb2}. UOTe is therefore not an outlier by field scale alone. It is unusual because a large $c$ axis metamagnetic field appears together with Van der Waals structure and metallic transport. 

The electronic structure makes this magnetic robustness consequential. Prior work established UOTe as an antiferromagnetic Dirac semimetal in which a Kondo-associated U $5f$ flat band hybridizes with Dirac-like conduction states, and in which the $\downarrow\uparrow\uparrow\downarrow$ magnetic structure preserves $\mathcal{P}\mathcal{T}$ symmetry in the bulk and odd-layer limit but breaks it in even layers \cite{UOTE_MAIN_PAPER}. Bilayer UOTe was consequently predicted to realize a fully compensated antiferromagnetic Chern-insulating state with nonzero Chern number despite zero net magnetization \cite{UOTE_MAIN_PAPER}. The present results add the magnetic constraint required for that picture: the compensated antiferromagnetic state is not fragile, but remains easy-axis locked, quasi-two-dimensional, and stable to large magnetic fields before entering a reconstructed high-field regime.

This combination is also relevant for antiferromagnetic spintronics, where zero-net-moment order, fast dynamics, spin-orbit coupling, and Berry-curvature effects can link magnetic symmetry to transport \cite{MacDonald2011AFMMetalSpintronics,Jungwirth2016AFMSpintronics,Baltz2018AFMSpintronics,CITE_TOPOLOGICAL_AFM_REVIEW}. Van der Waals antiferromagnets add thickness control and heterostructure integration, but many known examples are insulating or function mainly as tunnel barriers rather than active conducting channels \cite{Rahman2021VdWAFM2D}. UOTe combines exfoliability, compensated antiferromagnetism, $T_{\mathrm N}\approx150~\mathrm{K}$, a large spin-flip scale, metallic Dirac-like carriers, Kondo-associated hybridization, and even-layer $\mathcal{P}\mathcal{T}$ breaking. The central result is therefore not only that UOTe has a large spin-flip field, but that a quasi-two-dimensional Ising antiferromagnetic state survives inside a correlated, conducting, topological uranium material and reconstructs only through a broad, hysteretic metamagnetic regime with multiple internal scales.

\section{Conclusion}\label{sec:conclusion}

We have shown that UOTe realizes a \ quasi-two-dimensional Ising antiferromagnetic state whose rigidity persists from the paramagnetic regime into extreme magnetic fields. Susceptibility, neutron diffraction, and torque magnetometry establish a strongly uniaxial ordered state with moments locked to the crystallographic $c$ axis and critical behavior close to the two-dimensional Ising limit. Pulsed-field magnetization and angle-dependent PDO measurements then show that this state survives to giant magnetic fields, with the metamagnetic boundary controlled by the $c$-axis field component rather than by the total applied field. The field-induced reconstruction is broad, hysteretic, and internally structured, with no full saturation by 73~T, indicating that the high-field evolution is not a simple spin flip of rigid local moments.

These results place UOTe in a distinct regime among antiferromagnets. Its large metamagnetic spin flip field appears in a Van der Waals uranium material that is conducting, Kondo-interacting, and topologically nontrivial, rather than in a localized insulating magnet or a conventional three-dimensional oxide. The emergence of strong hysteresis below the same temperature scale associated with the U \(5f\) flat band further links the magnetic reconstruction to the correlated electronic background. UOTe therefore provides a rare platform in which reduced-dimensional Ising order, high-field antiferromagnetic stability, metallic transport, Kondo-associated hybridization, and symmetry-enabled topology coexist within a single compensated magnetic state.

\begin{acknowledgments}
We would like to thank Dmitry Chichinadze for fruitful discussions.
Research at Washington University was supported by the National Science Foundation (NSF) Division of Materials Research Award DMR-2236528. 
Z.R., S.G. and C.B. acknowledge the NRT LinQ, supported by the NSF under Grant No.
2152221.
A portion of this work was performed at the National High Magnetic Field Laboratory (NHMFL)  which is supported by National Science Foundation Cooperative Agreements No. DMR-1644779 and No. DMR-2128556 and the State of Florida and the Department of Energy (DOE). 
J.S. acknowledges support from the DOE BES program ”Science at 100 T”.
The work at Oak Ridge National Laboratory (ORNL) was supported by the U.S. DOE, Office of Science, Office of Basic Energy Sciences, Early Career Research Program Award KC0402020, under Contract DE-AC05-00OR22725. This research used resources at the High Flux Isotope Reactor, a DOE Office of Science User Facility operated by ORNL.

\end{acknowledgments}

\section*{Data Availability}

The data and analysis files that support the findings of this study are available from the corresponding author upon reasonable request. 

\bibliography{PRB_Main_Bibliography}

\end{document}


\title{Supplemental Material: Field-rigid Ising antiferromagnetism with giant spin-flip fields in the topological Kondo material UOTe}

\author{Zackary Rehfuss}
\affiliation{Department of Physics, Washington University in St. Louis, St. Louis, Missouri 63130, USA}

\author{Shannon Gould}
\affiliation{Department of Physics, Washington University in St. Louis, St. Louis, Missouri 63130, USA}

\author{Joanna Blawat}
\affiliation{National High Magnetic Field Laboratory, Los Alamos National Laboratory, Los Alamos, New Mexico 87545, USA}

\author{Christopher Broyles}
\affiliation{Department of Physics, Washington University in St. Louis, St. Louis, Missouri 63130, USA}

\author{George Xu}
\affiliation{Department of Physics, Washington University in St. Louis, St. Louis, Missouri 63130, USA}

\author{Yiqing Hao}
\affiliation{Neutron Scattering Division, Oak Ridge National Laboratory, Oak Ridge, TN 37831, USA}

\author{Huibo Cao}
\affiliation{Neutron Scattering Division, Oak Ridge National Laboratory, Oak Ridge, TN 37831, USA}

\author{Thao Dinh}
\affiliation{Department of Chemistry and Chemical Biology, Harvard University, Cambridge, MA 02138, USA}

\author{Suyang Xu}
\affiliation{Department of Chemistry and Chemical Biology, Harvard University, Cambridge, MA 02138, USA}

\author{Dave Graf}
\affiliation{National High Magnetic Field Laboratory, Florida State University, Tallahassee, Florida 32310, USA}

\author{John Singleton}
\affiliation{National High Magnetic Field Laboratory, Los Alamos National Laboratory, Los Alamos, New Mexico 87545, USA}

\author{Sheng Ran}
\affiliation{Department of Physics, Washington University in St. Louis, St. Louis, Missouri 63130, USA}

\date{\today}

\maketitle

\section{Neutron fitting details and fit-window dependence}
\label{sec:neutron_fitting_details}

To quantify the critical behavior of the antiferromagnetic order parameter, we fit the temperature dependence of the magnetic Bragg peak intensity to
\begin{equation}
I(T) \propto \left(1-\frac{T}{T_{\mathrm N}}\right)^{2\beta},
\label{eq:neutron_order_parameter_fit_supp}
\end{equation}
where \(I(T)\) is the integrated magnetic Bragg intensity, \(T_{\mathrm N}\) is the fitted ordering temperature, and \(\beta\) is the order-parameter critical exponent. The exponent appears as \(2\beta\) because the Bragg intensity is proportional to the square of the ordered magnetic moment.

A practical ambiguity in applying Eq.~\eqref{eq:neutron_order_parameter_fit_supp} is the choice of fitting window. The upper end of the fit is naturally set by the transition region near \(T_{\mathrm N}\), but the lower temperature bound is not uniquely defined. Including data too far below the transition can introduce noncritical temperature dependence into the extracted exponent, while using an overly narrow window makes the fit more sensitive to statistical scatter and to the fitted value of \(T_{\mathrm N}\). To test the robustness of the extracted exponent, we repeated the analysis over a series of windows with different lower bounds, \(T_{\mathrm{start}}\), while keeping the upper end of the fit near the transition.

\begin{figure}[t]
    \centering
    \maybeincludegraphics[width=\linewidth]{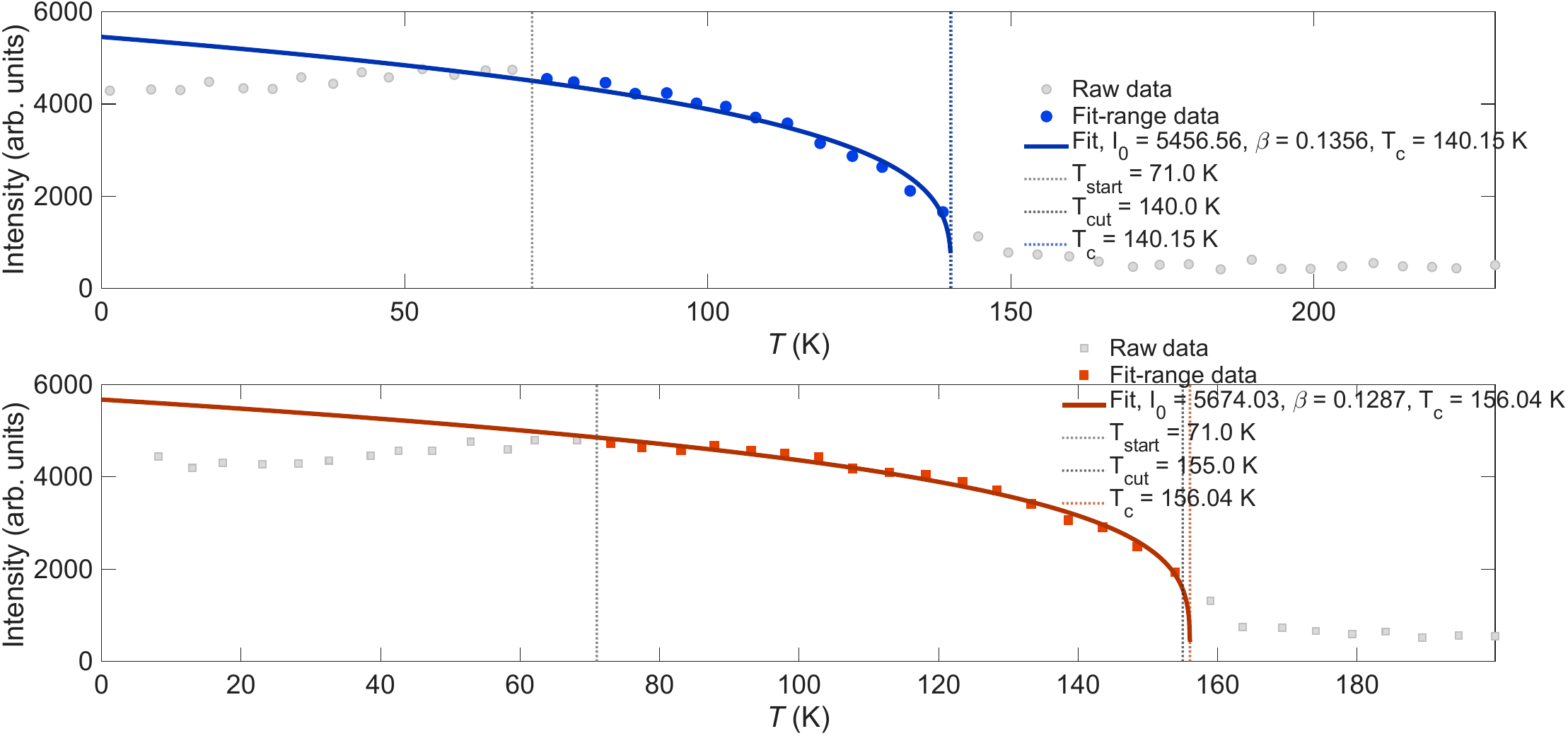}
    \caption{\textbf{Representative neutron critical-exponent fit.}
    Magnetic Bragg peak intensity fit to \(I(T)\propto (1-T/T_{\mathrm N})^{2\beta}\) over a representative fitting window beginning at \(T_{\mathrm{start}}=71~\mathrm{K}\) and extending to the transition region. This example illustrates the procedure used to extract \(T_{\mathrm N}\) and \(\beta\) from the temperature dependence of the magnetic order parameter.}
    \label{fig:neutron_sup_1}
\end{figure}

\begin{figure}[t]
    \centering
    \maybeincludegraphics[width=\linewidth]{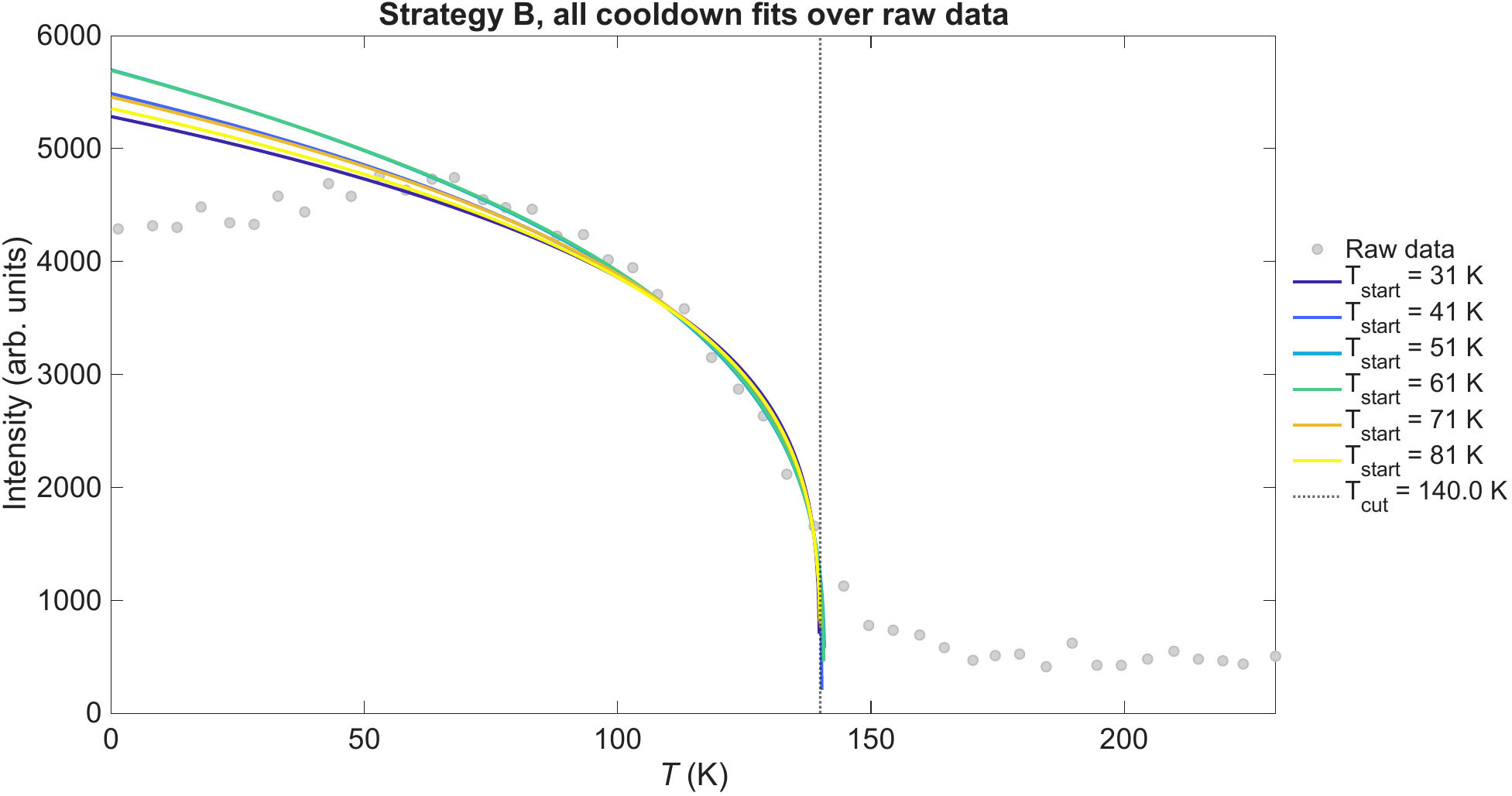}
    \caption{\textbf{Fit-window dependence for the cooldown neutron data.}
    The order-parameter fit was repeated for multiple choices of the lower fitting bound \(T_{\mathrm{start}}\). Each fit uses the same functional form, \(I(T)\propto (1-T/T_{\mathrm N})^{2\beta}\), while varying the temperature range included below the transition.}
    \label{fig:neutron_sup_2}
\end{figure}

\begin{figure}[t]
    \centering
    \maybeincludegraphics[width=\linewidth]{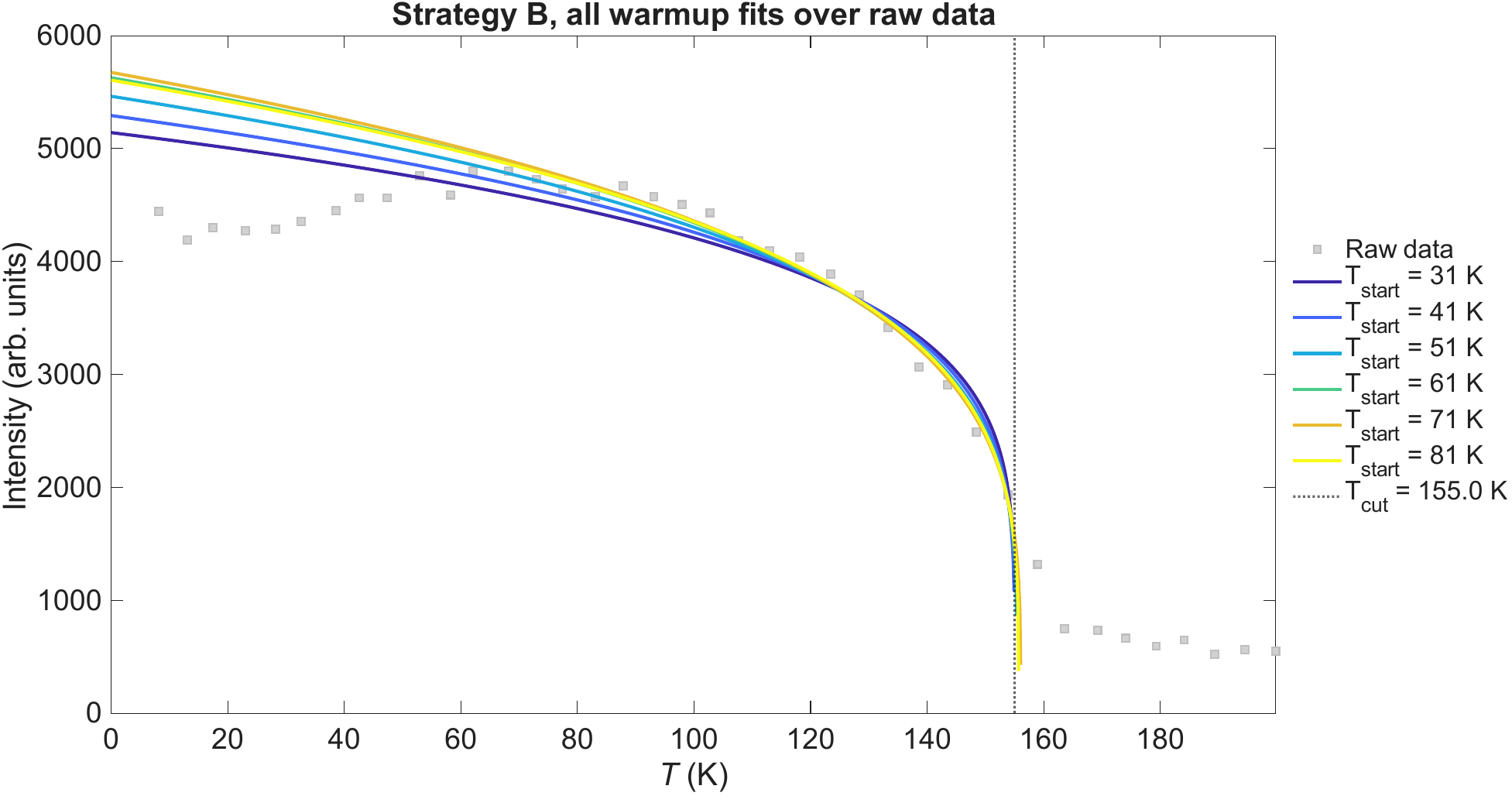}
    \caption{\textbf{Fit-window dependence for the warmup neutron data.}
    The same fitting procedure used for the cooldown data was applied to the warmup data set for multiple choices of \(T_{\mathrm{start}}\). The consistency of the fits provides an independent check on the robustness of the extracted order-parameter exponent.}
    \label{fig:neutron_sup_3}
\end{figure}

Figure~\ref{fig:neutron_sup_1} shows a representative fit over a window beginning at \(T_{\mathrm{start}}=71~\mathrm{K}\). The same procedure was then repeated for a set of lower-bound temperatures. The resulting fits for the cooldown and warmup data sets are shown in Figs.~\ref{fig:neutron_sup_2} and \ref{fig:neutron_sup_3}, respectively. This window-dependent analysis tests whether the extracted exponent is controlled by a single arbitrary choice of fitting range.

\begin{figure}[t]
    \centering
    \maybeincludegraphics[width=\linewidth]{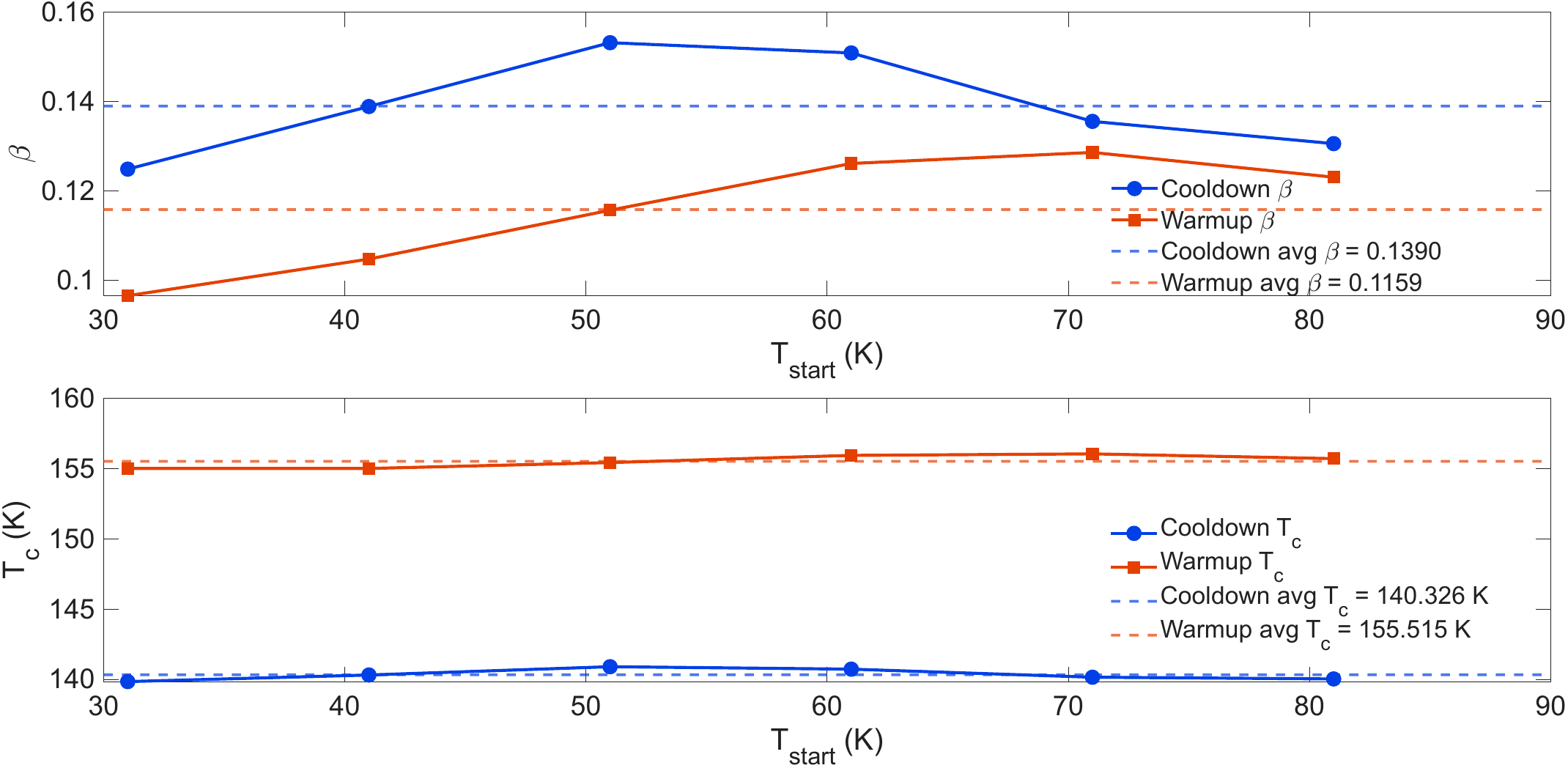}
    \caption{\textbf{Fit parameters as a function of lower fitting bound.}
    Extracted critical exponent \(\beta\) and ordering temperature \(T_{\mathrm N}\) as a function of \(T_{\mathrm{start}}\). The fitted \(\beta\) values remain close to the two-dimensional Ising value, \(\beta=1/8\), across the tested fitting windows. The window-averaged value is used to characterize the systematic uncertainty associated with the choice of fitting bound, while the best single fit is reported in the main text.}
    \label{fig:neutron_sup_4}
\end{figure}

The fitted values of \(\beta\) and \(T_{\mathrm N}\) are summarized as a function of \(T_{\mathrm{start}}\) in Fig.~\ref{fig:neutron_sup_4}. Across the tested windows, \(\beta\) remains close to the two-dimensional Ising value, \(\beta=1/8\). The fitted parameters show only a weak systematic dependence on the lower temperature bound. Extending the fit to lower temperatures slightly reduces the extracted \(\beta\), consistent with the inclusion of data outside the narrowest critical regime. We therefore report both the best single fit used in the main text and the average value obtained over the tested fitting windows. This procedure shows that the assignment of nearly two-dimensional Ising critical behavior does not depend sensitively on one arbitrary fitting bound.

\section{VSM data used to normalize magnetization and torque}
\label{sec:vsm_normalization}

\begin{figure}[t]
    \centering
    \maybeincludegraphics[width=\linewidth]{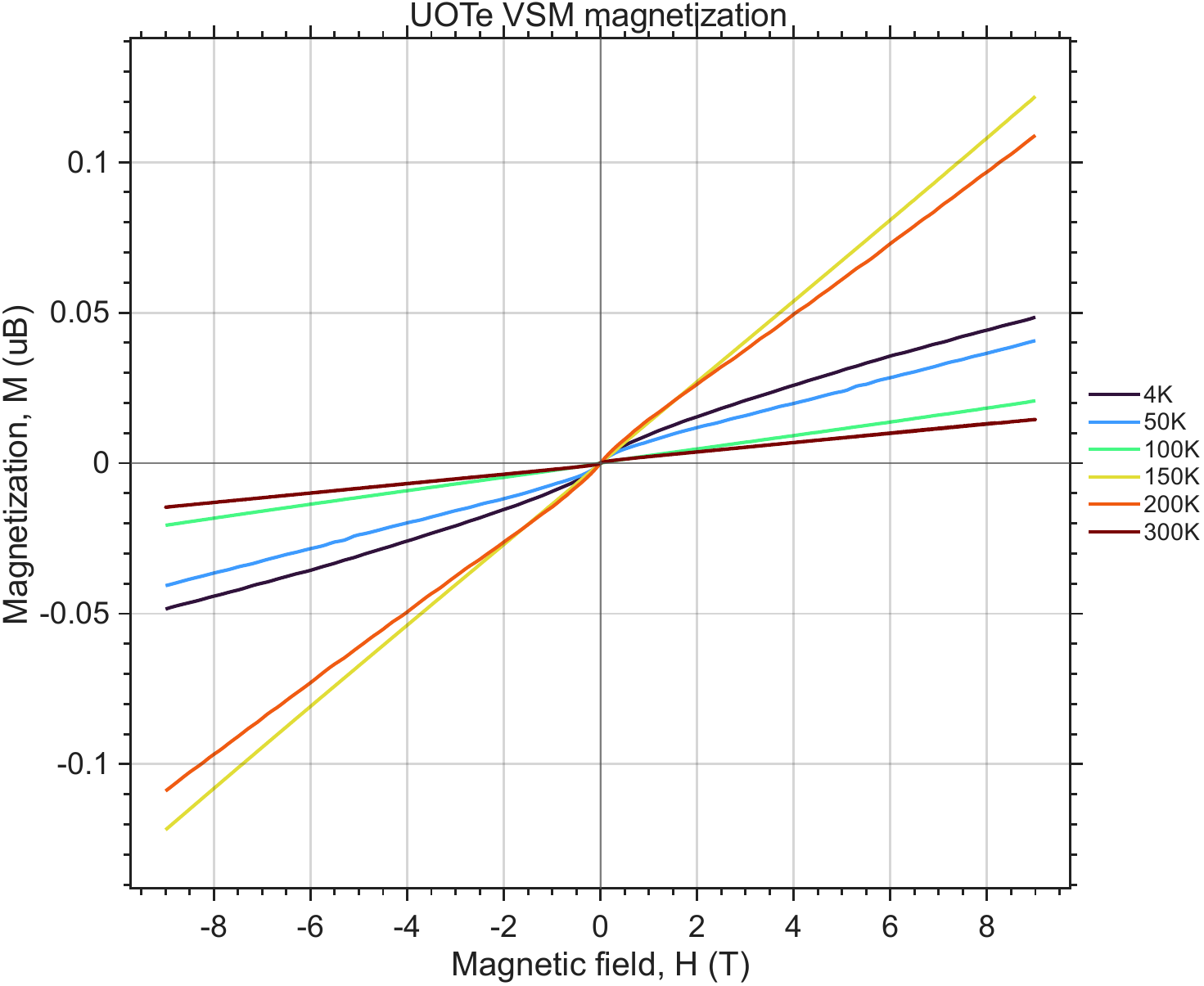}
    \caption{\textbf{VSM normalization data.}
    Vibrating sample magnetometry data used to normalize the pulsed-field magnetization and torque measurements. The data span \(4~\mathrm{K}\) to \(300~\mathrm{K}\), and provide the absolute magnetization scale used to express the field-dependent measurements in units of \(\mu_{\mathrm B}\) per uranium atom.}
    \label{fig:vsm_sup}
\end{figure}

Vibrating sample magnetometry measurements were used to place the pulsed-field magnetization and torque data on an absolute moment scale. The VSM measurements span temperatures from \(4~\mathrm{K}\) to \(300~\mathrm{K}\), providing an independent low-field magnetization reference for converting the pulsed-field magnetization and torque responses into units of \(\mu_{\mathrm B}\) per uranium atom. The VSM data used for this normalization are shown in Fig.~\ref{fig:vsm_sup}.

\section{Torque magnetometry data}
\label{sec:torque_florida}

\begin{figure}[t]
    \centering
    \maybeincludegraphics[width=\linewidth]{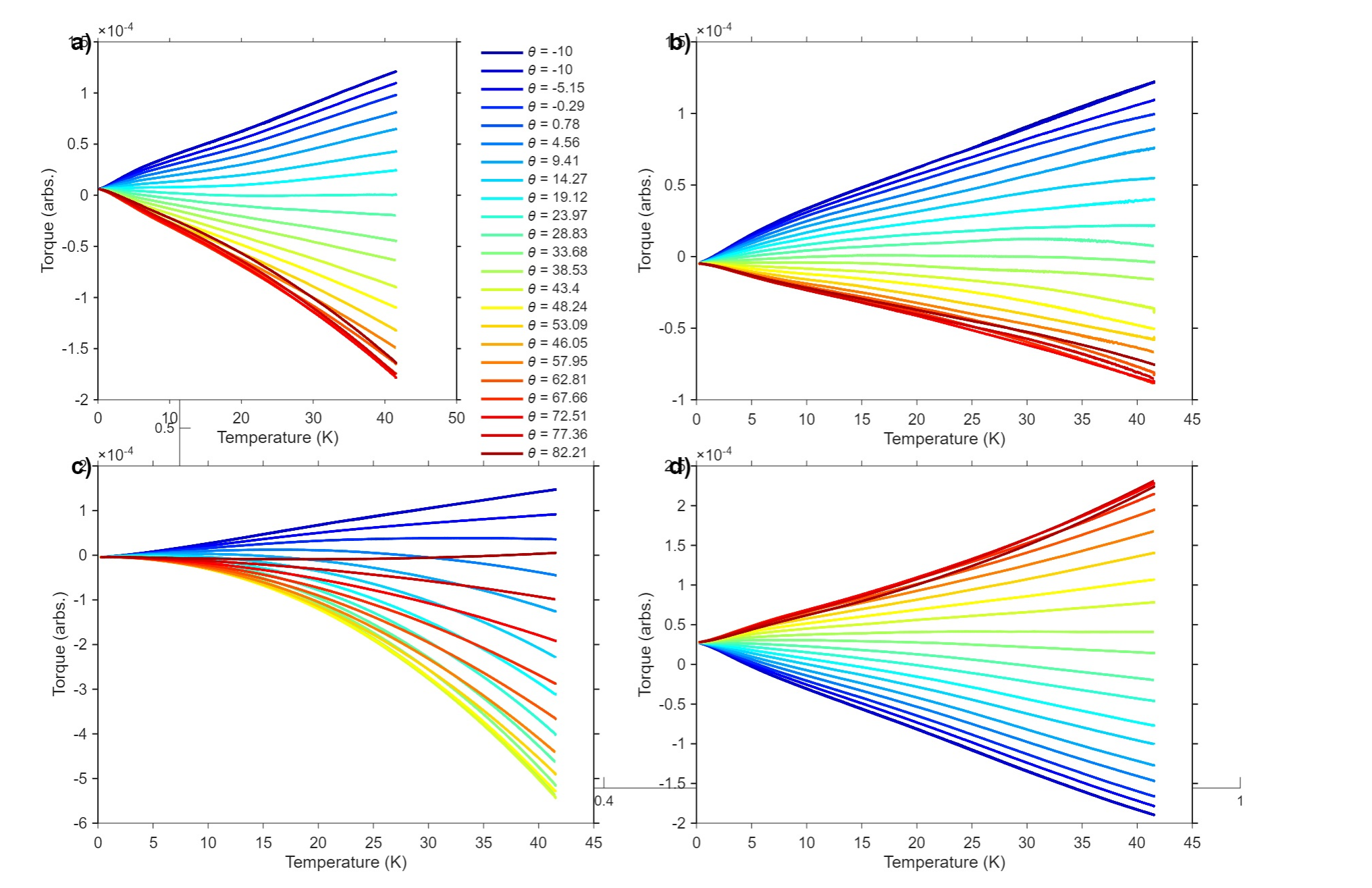}
    \caption{\textbf{Raw torque magnetometry data.}
    Torque as a function of magnetic field for four UOTe samples measured in the same experimental run. Differences among the traces arise primarily from sample mounting geometry and angular offsets between measurement windows, while the overall field dependence remains consistent across samples.}
    \label{fig:Torque_Raw_Florida}
\end{figure}

Magnetic torque was measured as a function of field and angle on several UOTe samples to test whether the ordered state retains a simple uniaxial response below the high-field transition. The raw torque data are shown in Fig.~\ref{fig:Torque_Raw_Florida}. These measurements were collected from four UOTe samples in the same experimental run. Because the samples were mounted with different orientations in separate measurement windows, the raw traces differ in sign and angular phase, while preserving the same underlying field dependence.

To analyze the angular response, the full torque data set was combined across field and angle to extract representative angular cuts at fixed field values. These cuts are shown in Fig.~\ref{fig:Torque_Fit_AngleSlices}. Across the accessible field range, the data are well described by the angular dependence expected for a uniaxial two-sublattice antiferromagnet\cite{Kittel1951AFMR,Nagamiya1955Antiferromagnetism},
\begin{equation}
\tau \propto \Delta\chi\,H^{2}\sin(2\theta),
\label{eq:torque_supp}
\end{equation}
where \(\Delta\chi=\chi_c-\chi_{ab}\) is the susceptibility anisotropy. The torque amplitude therefore tracks the difference between the \(c\)-axis and in-plane magnetic responses. The observed \(\sin(2\theta)\) dependence shows that the ordered state remains well described by a uniaxial antiferromagnet below the high-field transition.

\begin{figure}[t]
    \centering
    \maybeincludegraphics[width=\linewidth]{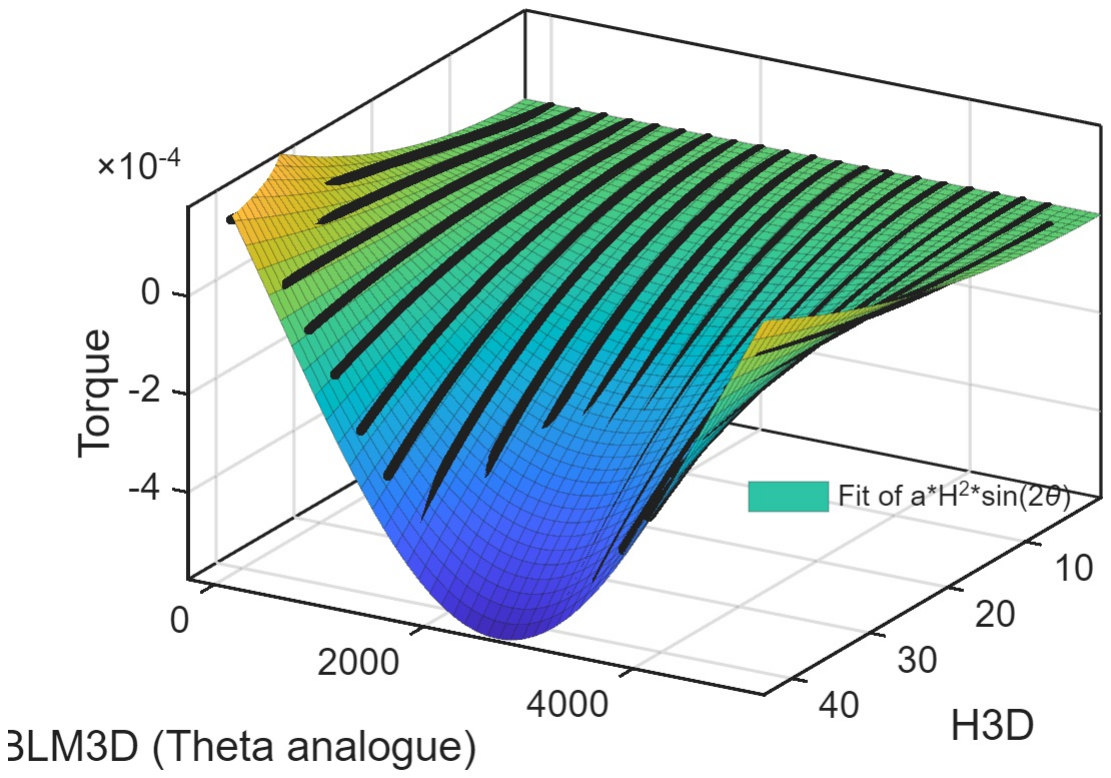}
    \caption{\textbf{Global torque fit.}
    Three-dimensional representation of the torque as a function of magnetic field and angle. The overall field and angular dependence follows the expected \(\sin(2\theta)H^2\) scaling for a uniaxial two-sublattice antiferromagnet.}
    \label{fig:Torque_Fit_3D_Surface}
\end{figure}

A complementary view of the same data is shown in Fig.~\ref{fig:Torque_Fit_3D_Surface}, where the torque is plotted as a function of field and angle. The full data set follows the expected \(\sin(2\theta)H^2\) scaling with good overall agreement, providing a global visualization of the uniaxial two-sublattice response discussed in the main text.

\begin{figure}[t]
    \centering
    \maybeincludegraphics[width=\linewidth]{Torque_Sup_2.pdf}
    \caption{\textbf{Angular cuts through the torque data.}
    Representative angular cuts extracted from the torque data at fixed magnetic fields. Solid lines are fits to the \(\sin(2\theta)\) angular dependence expected for a uniaxial two-sublattice antiferromagnet, \(\tau \propto \Delta\chi H^2\sin(2\theta)\). The agreement shows that the angular response remains consistent with a simple uniaxial antiferromagnetic state throughout the measured field range.}
    \label{fig:Torque_Fit_AngleSlices}
\end{figure}

\section{TDO and magnetoresistance data from Tallahassee}
\label{sec:tdo_mr_florida}

Additional tunnel-diode oscillator (TDO) and magnetoresistance measurements were performed in Tallahassee in static magnetic fields up to \(\mu_0H=41.5~\mathrm{T}\). These measurements were not used as primary data in the main text, but they provide an additional check on the absence of field-induced transitions below the spin-flip regime identified by pulsed-field magnetization and PDO measurements. In both the TDO and magnetoresistance data, no reproducible transition-like feature is observed below \(41.5~\mathrm{T}\) for the measured field orientations.

\begin{figure}[t]
\centering
\maybeincludegraphics[width=\linewidth]{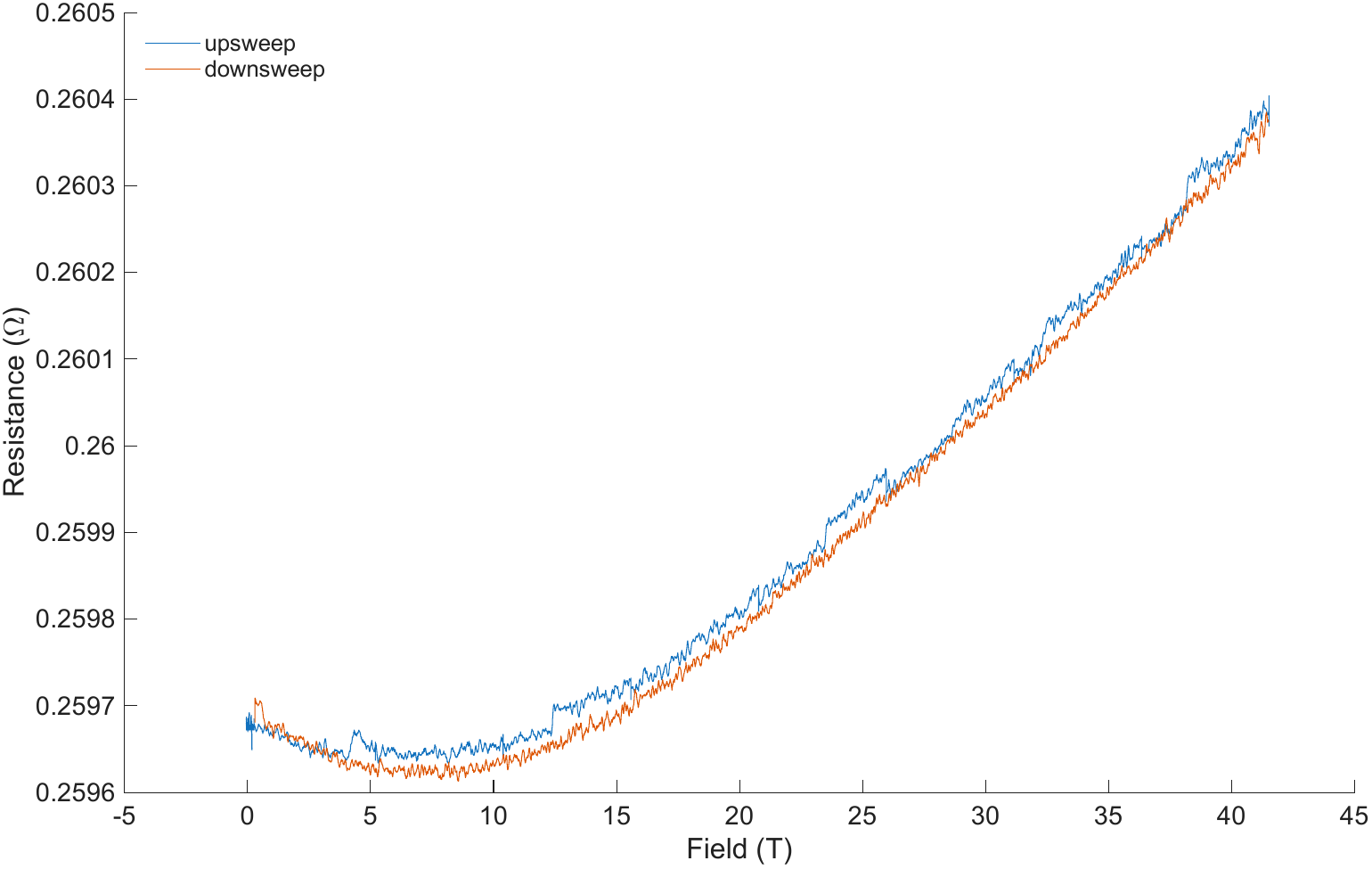}
\caption{\textbf{Magnetoresistance sweep comparison below the high-field transition.}
Representative magnetoresistance data measured on increasing and decreasing field sweeps at the same temperature and field angle. No measurable hysteresis is observed below \(\mu_0H=41.5~\mathrm{T}\), consistent with the absence of a field-induced transition in this field range.}
\label{fig:MR_Sup_1}
\end{figure}

Figure~\ref{fig:MR_Sup_1} shows representative magnetoresistance data measured on increasing and decreasing field sweeps at the same temperature and field angle. The up-sweep and down-sweep traces overlap within the experimental scatter, indicating that no measurable hysteresis occurs below \(41.5~\mathrm{T}\). This is consistent with the pulsed-field magnetization results, where strong hysteresis appears only in the high-field metamagnetic regime near and above the spin-flip onset.

\begin{figure}[t]
\centering
\maybeincludegraphics[width=\linewidth]{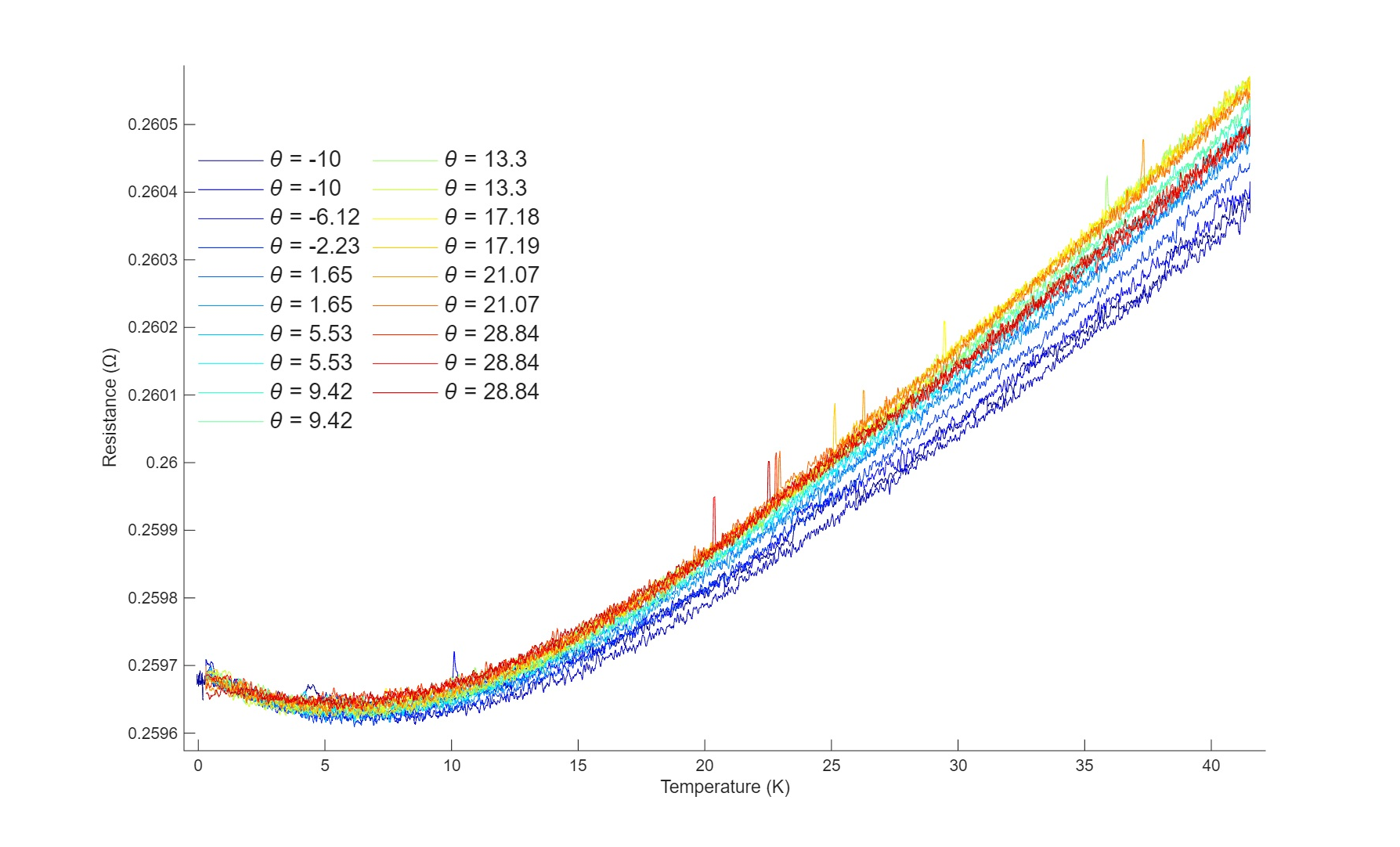}
\caption{\textbf{Angle-dependent magnetoresistance below the spin-flip field.}
Magnetoresistance of UOTe measured up to \(\mu_0H=41.5~\mathrm{T}\) for different field angles relative to the crystallographic \(c\) axis. The response is weakly angle dependent and shows no transition-like feature below the high-field spin-flip regime. The slightly larger deviations at high field occur as the measurement approaches, but remains below, the metamagnetic onset observed near \(50\)--\(55~\mathrm{T}\).}
\label{fig:MR_Sup_2}
\end{figure}

Figure~\ref{fig:MR_Sup_2} shows magnetoresistance measured for several field orientations relative to the crystallographic \(c\) axis. The magnetoresistance exhibits only weak angular dependence over most of the measured field range. A modest angular variation develops at the highest fields, as the applied field approaches the lower edge of the metamagnetic regime, but no sharp anomaly or discontinuity is observed. These data support the conclusion that fields below \(41.5~\mathrm{T}\) do not destabilize the antiferromagnetic state, even when the field is rotated with respect to the ordered moment direction.

\begin{figure}[t]
\centering
\maybeincludegraphics[width=\linewidth]{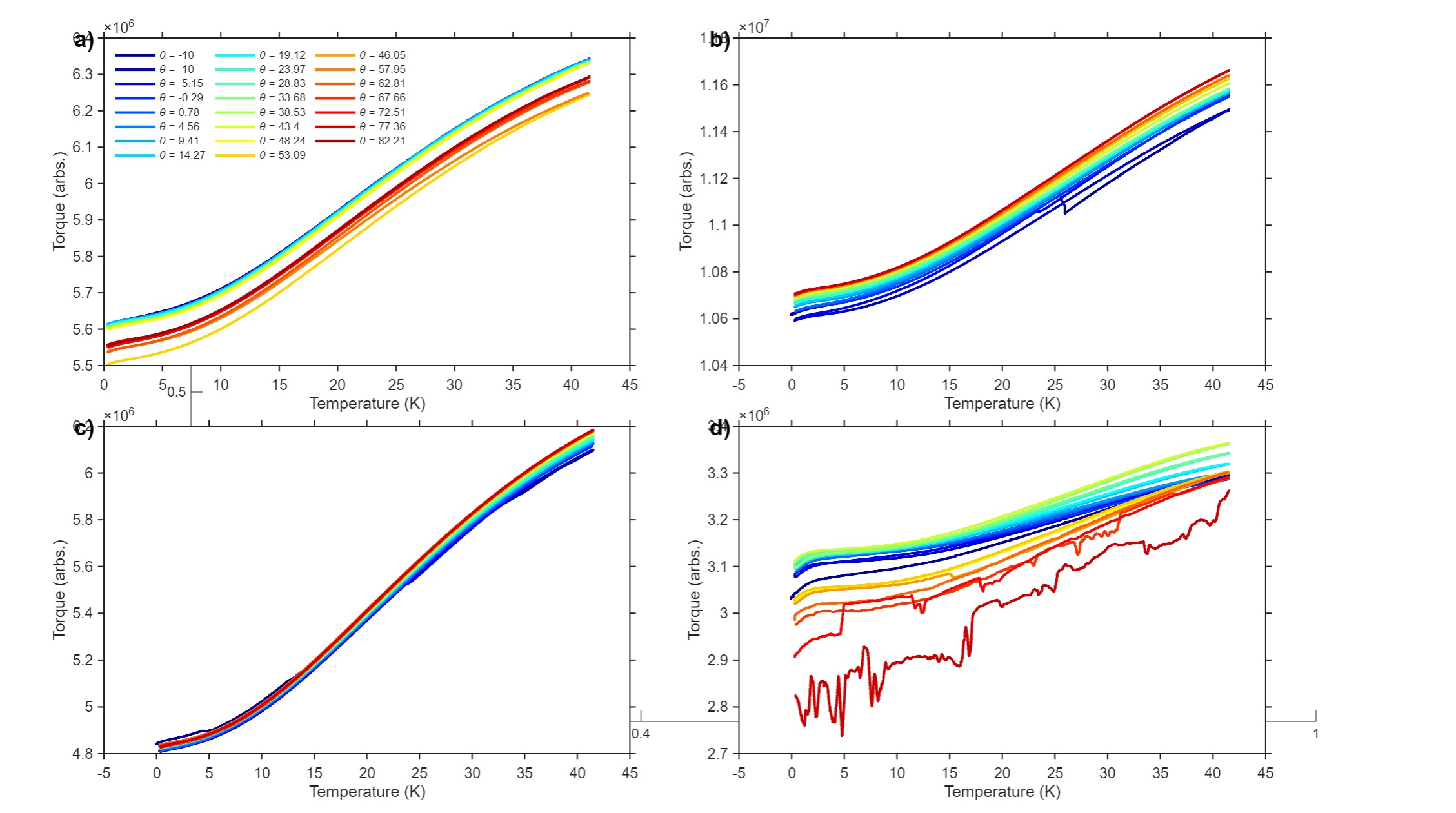}
\caption{\textbf{Raw TDO response below the high-field transition.}
Raw tunnel-diode oscillator data measured on four UOTe samples in static magnetic fields up to \(\mu_0H=41.5~\mathrm{T}\). No reproducible transition-like feature is observed below the spin-flip field. Non-reproducible noise features arise from TDO drive-power instability rather than from the sample response.}
\label{fig:TDO_Sup}
\end{figure}

The corresponding raw TDO data are shown in Fig.~\ref{fig:TDO_Sup}. The four panels show data collected from four UOTe samples measured in the same high-field environment. Apart from non-reproducible features associated with TDO drive-power instability, the traces do not show reproducible anomalies below \(41.5~\mathrm{T}\). The absence of sharp TDO features in this field range agrees with the magnetoresistance data and with the main high-field results, which place the spin-flip onset at substantially higher fields.

Together, the Tallahassee magnetoresistance and TDO measurements provide a control data set below the high-field transition. The lack of hysteresis, the weak angular magnetoresistance response, and the absence of reproducible TDO anomalies all support the conclusion that the antiferromagnetic state remains stable up to at least \(41.5~\mathrm{T}\), below the spin-flip onset identified near \(50\)--\(55~\mathrm{T}\).

\section{Magnetization analysis}
\label{sec:magnetization_analysis}

\begin{figure}[t]
\centering
\maybeincludegraphics[width=\linewidth]{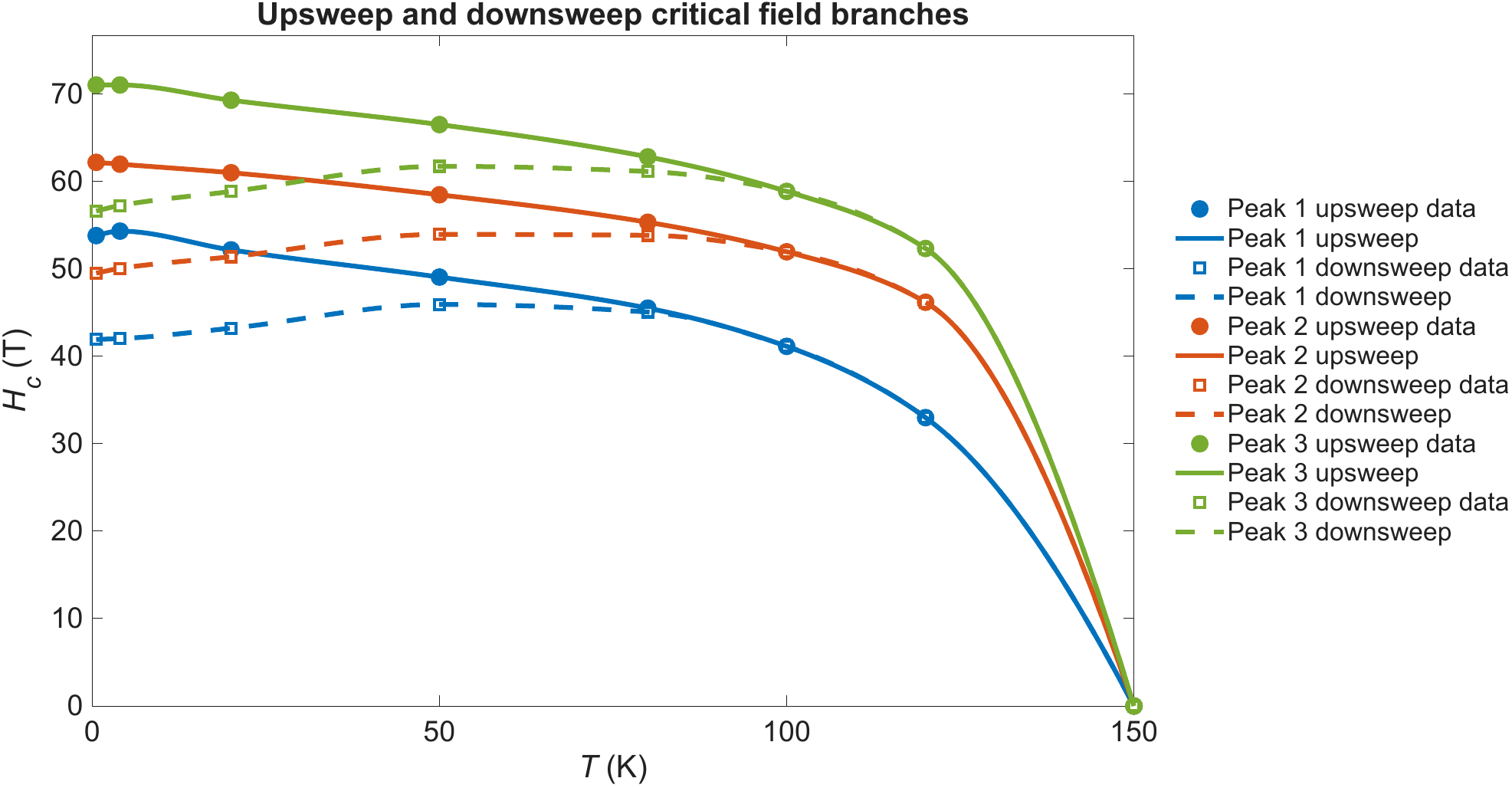}
\caption{\textbf{Derivative features tracked in the pulsed-field magnetization.}
Field derivatives of the magnetization showing the three characteristic features tracked through the high-field metamagnetic regime. Both up-sweep and down-sweep traces are shown. These derivative peaks define the field scales used to follow the onset, evolution, and upper portion of the broad spin-flip regime.}
\label{fig:Mag_Sup_1}
\end{figure}

The pulsed-field magnetization data were analyzed by tracking reproducible derivative features and by converting the raw pickup-coil signal to an absolute magnetization scale. Figure~\ref{fig:Mag_Sup_1} summarizes the derivative analysis used to follow the metamagnetic regime. Three derivative features were tracked on both the up-sweep and down-sweep traces. These features correspond to the onset, internal structure, and upper portion of the broad metamagnetic response discussed in the main text. The extracted peak positions are used in the field-temperature phase diagram, while the supplemental figure shows the full tracking procedure explicitly.

\begin{figure}[t]
\centering
\maybeincludegraphics[width=\linewidth]{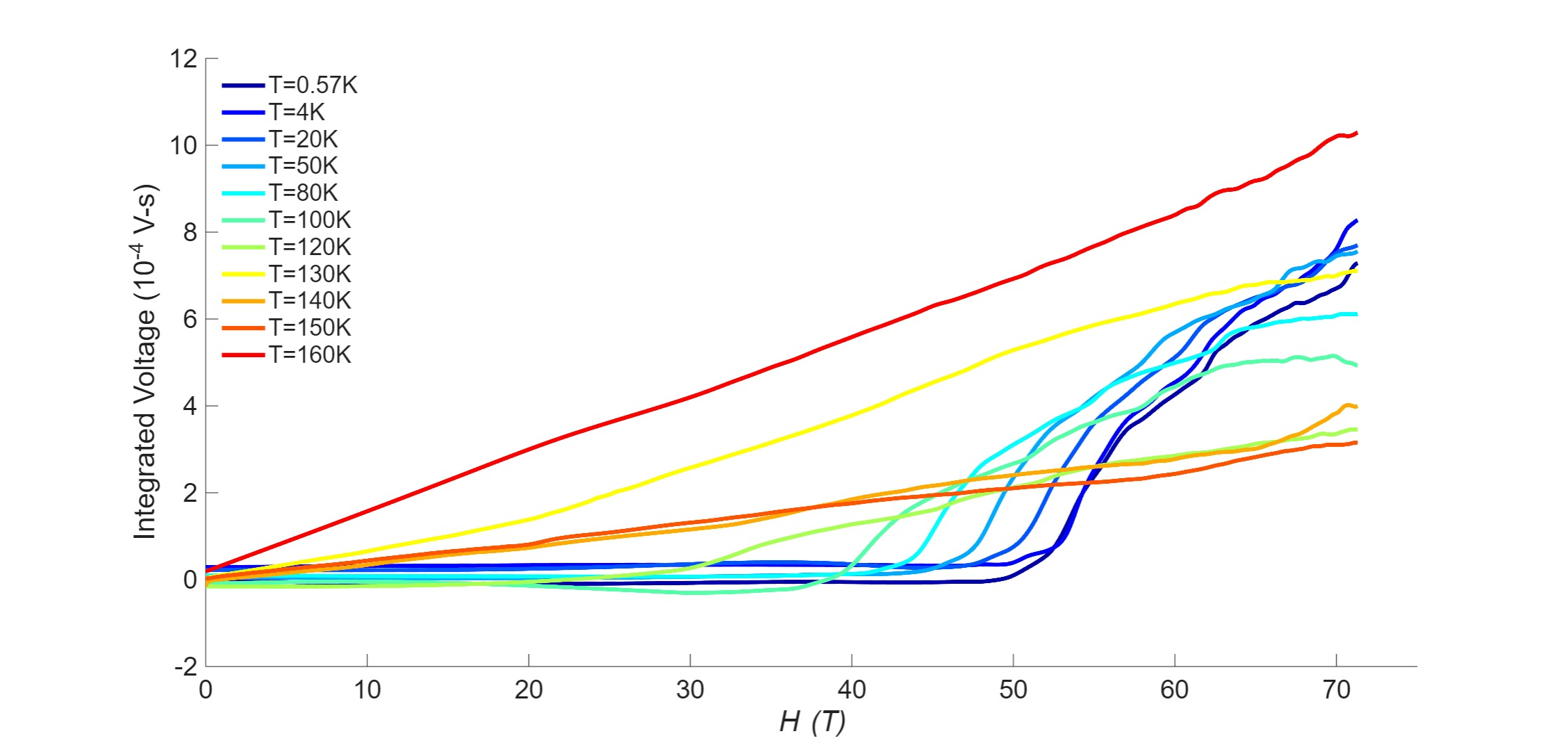}
\caption{\textbf{Raw integrated pickup-coil signal.}
Representative up-sweep pulsed-field magnetization data shown in raw integrated-voltage units. The integrated signal is proportional to magnetization but is not yet converted to \(\mu_{\mathrm B}\) per uranium atom and contains small offsets that must be corrected during normalization.}
\label{fig:Mag_Sup_2}
\end{figure}

The raw pulsed-field signal is shown in Fig.~\ref{fig:Mag_Sup_2}. The pickup-coil voltage was integrated to obtain a signal proportional to magnetization, but the resulting quantity is not initially in absolute magnetic units and contains run-dependent offsets. The up-sweep traces are shown as representative raw data. The down-sweep traces show the same overall behavior, but are omitted from this panel for clarity. This integrated-voltage data set is the starting point for the normalization procedure described below.

\begin{figure}[t]
\centering
\maybeincludegraphics[width=\linewidth]{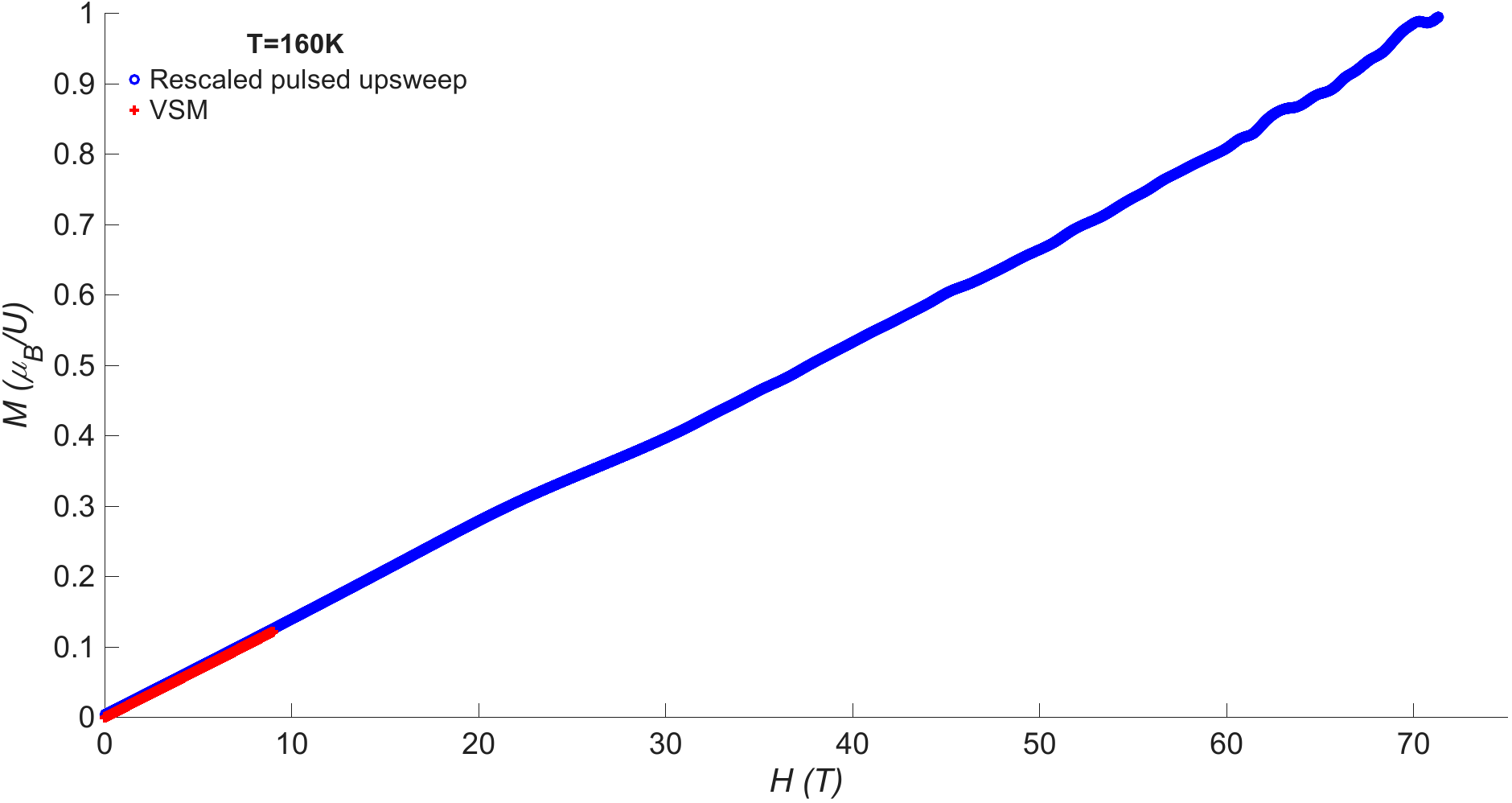}
\caption{\textbf{VSM-based normalization of the pulsed-field magnetization.}
Comparison between the VSM magnetization data and the \(160~\mathrm{K}\) pulsed-field magnetization trace after slope matching in the low-field linear regime. This procedure determines the multiplicative scale factor that converts the integrated pickup-coil signal into \(\mu_{\mathrm B}\) per uranium atom. The same procedure was applied to the high-temperature traces with linear low-field response.}
\label{fig:Mag_Sup_3}
\end{figure}

To convert the integrated voltage into absolute magnetization units, we used vibrating sample magnetometry as an independent calibration reference. Figure~\ref{fig:Mag_Sup_3} shows the calibration procedure for the \(160~\mathrm{K}\) pulsed-field trace. At this temperature, the low-field magnetization is approximately linear in field, allowing the slope of the integrated pulsed-field signal to be matched to the VSM magnetization. The same slope-matching procedure was applied to all traces with a sufficiently linear low-field response, corresponding to temperatures above \(100~\mathrm{K}\). The resulting scale factors were nearly identical across the fitted traces, indicating that the pickup-coil calibration was stable between runs. These scale factors were then used to transform the full pulsed-field data set into units of \(\mu_{\mathrm B}\) per uranium atom.

After determining the multiplicative normalization, a small additive offset was applied to place the magnetization on the correct zero-field scale. The integrated-voltage traces contain offsets because the low-field signal is small compared with the high-field response and because numerical integration of pickup-coil voltages can introduce baseline shifts. To correct this, we combined the VSM, torque magnetometry, and pulsed-field magnetization data. The VSM data determine the absolute low-field magnetization scale, while the torque data provide an independent constraint on the smooth low-field response up to \(41.5~\mathrm{T}\), below the metamagnetic transition. This low-field behavior was then used to determine the intercept needed to align the pulsed-field magnetization before the high-field jump.

\begin{figure}[t]
\centering
\maybeincludegraphics[width=\linewidth]{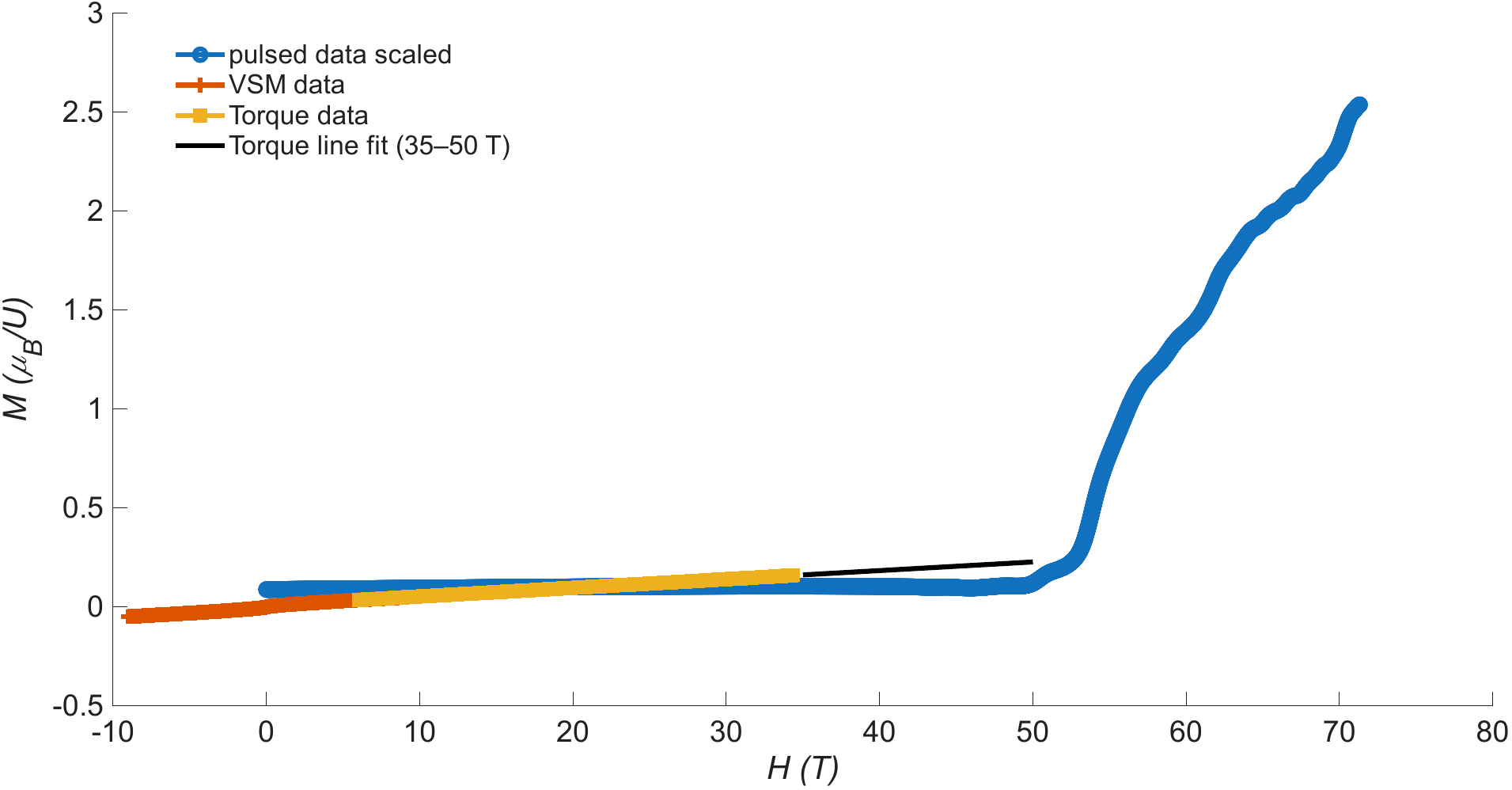}
\caption{\textbf{Offset correction using VSM, torque, and pulsed-field magnetization.}
Procedure used to determine the additive offset in the pulsed-field magnetization data after multiplicative normalization. The VSM data provide the absolute low-field magnetization scale, while the torque data constrain the smooth magnetization response below \(41.5~\mathrm{T}\). These constraints are combined with the pulsed-field traces to set the zero-field offset and align the low-field response before the metamagnetic transition.}
\label{fig:Mag_Sup_4}
\end{figure}

Operationally, the raw pulsed-field signal was transformed according to
\begin{equation}
M(H,T)=m\,V_{\mathrm{int}}(H,T)+b,
\label{eq:mag_norm_supp}
\end{equation}
where \(V_{\mathrm{int}}\) is the integrated pickup-coil voltage, \(m\) is the multiplicative calibration factor determined from the VSM comparison, and \(b\) is the additive offset determined by matching the low-field magnetization to the combined VSM and torque constraints. The result of this procedure is shown in Fig.~\ref{fig:Mag_Sup_4}. This calibration converts the pickup-coil voltage into absolute magnetization units while enforcing the physical condition that the measured magnetization is zero at \(H=0\).

\begin{figure}[t]
\centering
\maybeincludegraphics[width=\linewidth]{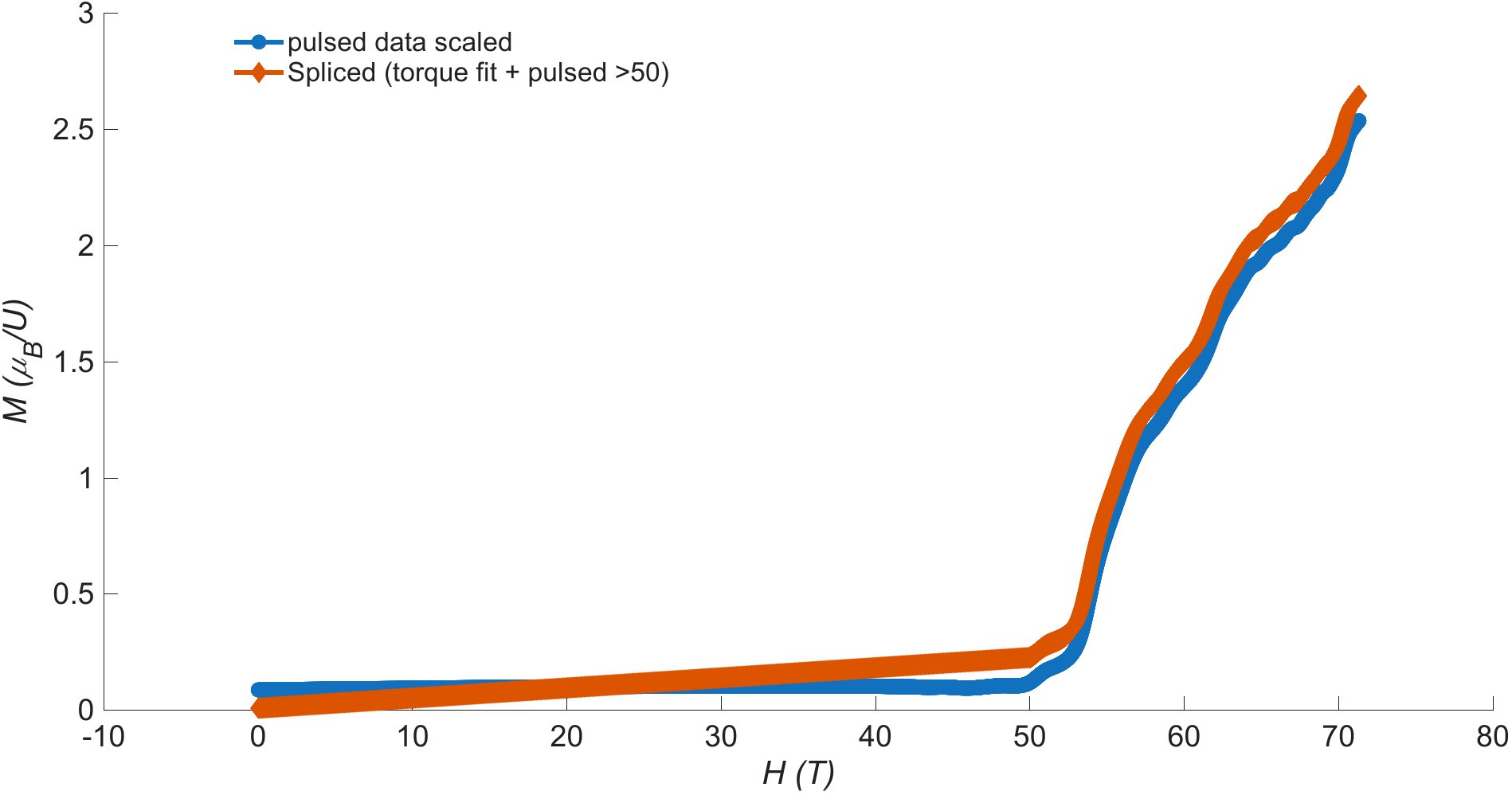}
\caption{\textbf{Effect of offset correction on the normalized magnetization.}
Comparison between the normalized pulsed-field magnetization before and after applying the additive offset correction. The correction mainly improves the low-field baseline, where the integrated pickup-coil signal is small. The high-field response and the derivative features used to determine the spin-flip and metamagnetic field scales are unchanged.}
\label{fig:Mag_Sup_5}
\end{figure}

Figure~\ref{fig:Mag_Sup_5} compares the normalized pulsed-field data before and after applying the offset correction. The correction improves the low-field baseline, where the raw pickup-coil signal is smallest and most sensitive to integration offsets. The high-field magnetization curve is otherwise essentially unchanged. In particular, the offset correction does not shift the spin-flip onset field, the derivative peak positions, the hysteresis width, or any of the field scales used in the phase diagram. The calibration therefore affects the absolute vertical scale and low-field baseline, but not the extracted transition fields.

Together, this analysis converts the raw pickup-coil voltage into an absolute magnetization scale by combining independently calibrated VSM data, the smooth low-field torque response, and the pulsed-field integrated-voltage traces. The procedure determines the vertical normalization of the magnetization curves but does not affect the field positions of the metamagnetic anomalies reported in the main text.

\section{Extended PDO data}
\label{sec:pdo_extended}

\begin{figure}[t]
\centering
\maybeincludegraphics[width=\linewidth]{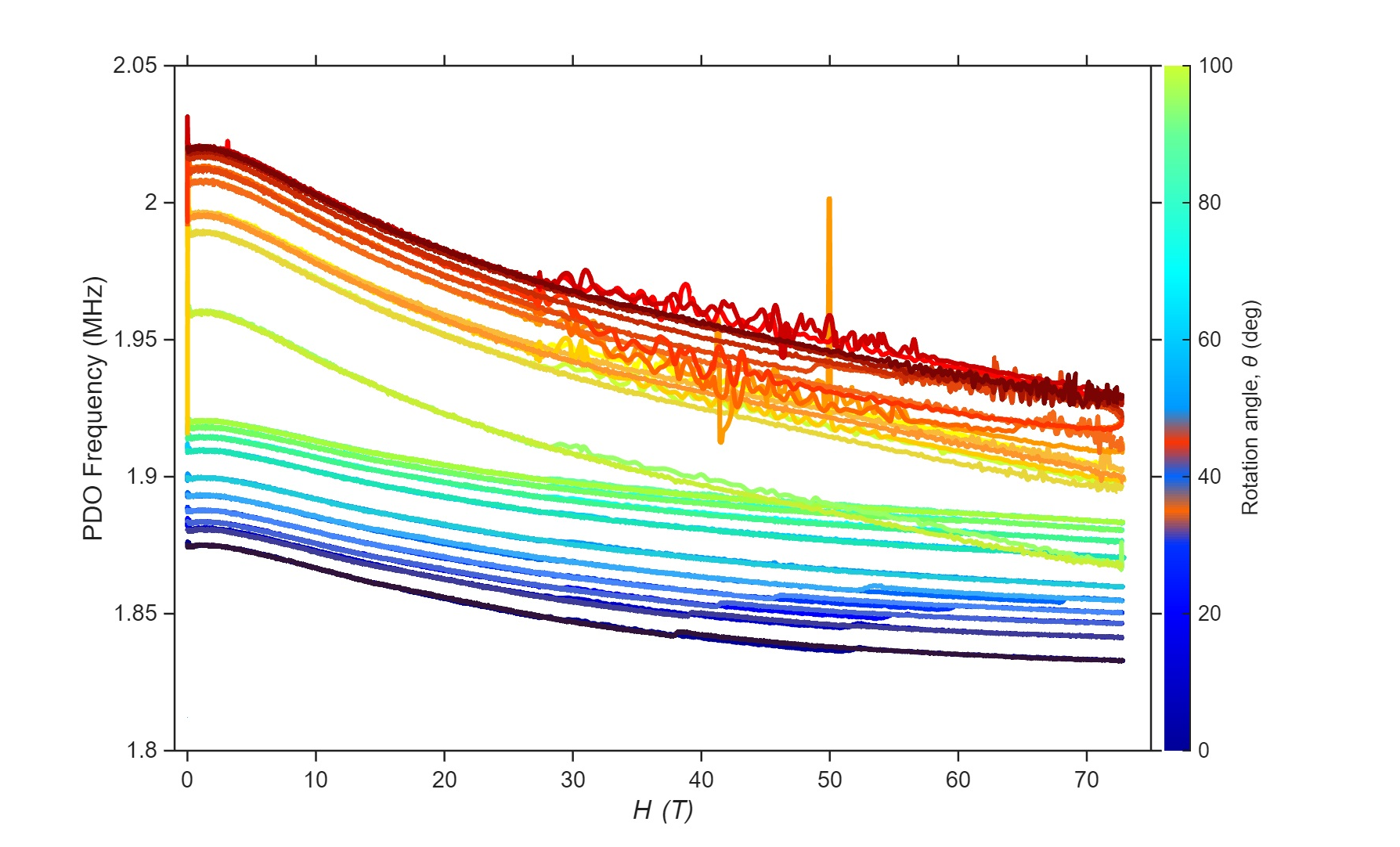}
\caption{\textbf{Extended PDO data set.}
Raw proximity detector oscillator data measured over an extended range of field angles and temperatures. The color scale indicates the field angle for the continuous angle-dependent measurements. The two red markers on the color bar indicate the fixed angles used for additional temperature-dependent measurements, corresponding to the red and orange traces. The low-angle traces resolve the high-field anomaly used in the main text to determine the angular dependence of the spin-flip field. Additional high-angle and near-\(ab\)-plane measurements show no reproducible transition-like features within the accessible field range. The higher-angle and temperature-dependent traces were collected in separate runs with different PDO operating conditions, which produced increased instrumental noise, but no additional field-induced anomaly is resolved.}
\label{fig:PDO_Sup}
\end{figure}

Additional proximity detector oscillator (PDO) measurements were performed over a wider range of field angles and temperatures than the subset shown in the main text. These data are shown in Fig.~\ref{fig:PDO_Sup}. The main manuscript focuses on the low-angle data because those traces resolve the high-field anomaly clearly and allow the angular dependence of the critical field to be extracted. The extended data are included here to report the full set of measurements.

The color scale in Fig.~\ref{fig:PDO_Sup} indicates the field angle for the continuous angle-dependent measurements. Two red markers on the color bar identify the fixed angles at which additional temperature-dependent PDO measurements were performed. The corresponding red and orange traces therefore represent temperature sweeps at those fixed angles, rather than additional points in the continuous angular series.

The higher-angle and temperature-dependent traces were collected in separate runs under different PDO operating conditions, including changes to the oscillator settings. As a result, these traces contain more instrumental noise than the low-angle data used for the quantitative analysis. Nevertheless, no reproducible transition-like anomaly is observed in the high-angle data or in measurements with the field close to the \(ab\) plane within the accessible field range. This is consistent with the main-text conclusion that the metamagnetic instability is controlled by the field component along the crystallographic \(c\) axis. As the field is rotated toward the \(ab\) plane, the required total field is pushed beyond the available field window, and no independent in-plane transition scale is resolved.

\bibliography{PRB_Supplement_Bibliography}